\documentclass[10pt,twocolumn,twoside]{IEEEtran}

\ifCLASSINFOpdf
  
\else
  
\fi

\usepackage{xcolor}
\newtheorem{assm}{Assumption}

\newtheorem{rem}{Remark}
\newtheorem{defn}{Definition}
\newtheorem{lem}{Lemma}
\newtheorem{prop}{Proposition}
\newtheorem{cor}{Corollary}
\newtheorem{thm}{Theorem}

\usepackage{amsmath}
\usepackage{amstext}
\usepackage{amssymb}
\usepackage{tikz,overpic}
\usepackage[mathscr]{euscript}
\usepackage{color}
\usepackage{amsfonts}
\usepackage{enumitem}
\usepackage{amsmath}

\usepackage{stfloats}

\usepackage{epstopdf}
\makeatletter
\newcommand*{\rom}[1]{\expandafter\@slowromancap\romannumeral #1@}
\makeatother
\usetikzlibrary{automata}
\usepackage{graphicx}      % include this line if your document contains figures

\usepackage{xparse}% http://ctan.org/pkg/xparse
\DeclareMathAlphabet\mathbfcal{OMS}{cmsy}{b}{n}
\usepackage{array}
\usepackage{csquotes}

\usepackage{cuted}
\usepackage{mathtools}

\DeclarePairedDelimiter{\norm}{\lVert}{\rVert}
\usepackage{xcolor}

\usepackage[space]{cite}
\usepackage{comment}

\newcommand{\seb}[1]{%
{\leavevmode\color{black}#1}%
}

\newcommand{\sebcancel}[1]{%
{\leavevmode\color{black}#1}%
}

\newcommand{\phil}[1]{%
{\leavevmode\color{black}#1}%
}

\newcommand{\philnew}[1]{%
{\leavevmode\color{black}#1}%
}

\DeclareMathOperator{\diag}{diag}
\begin{document}
\title{Analysis and Distributed Control of\\ Periodic  Epidemic Processes 
% over Periodic  Time-Varying Networks 
% in Discrete-Time
}
\author{Sebin~Gracy,%~\IEEEmembership{Student Member,~IEEE,}
        ~Philip.~E.~Par\'e,~Henrik~Sandberg%~\IEEEmembership{Member,~IEEE,}
        ~and~Karl~Henrik~Johansson%~\IEEEmembership{Member,~IEEE} %<-this % stops a space
\thanks{The authors are with the Division of Decision and Control Systems, School of Electrical Engineering and Computer Science, KTH Royal Institute of Technology, Stockholm, Sweden. E-mails: gracy@kth.se, philipar@kth.se, hsan@kth.se, kallej@kth.se. This work
was supported in part by the Knut and Alice Wallenberg Foundation and
 the Swedish Research Council, VR 2016-00861. } }

\maketitle
\thispagestyle{plain} %--to make title page number less
\pagestyle{plain}    % -- to make other pages number less.
\begin{abstract} 
This paper studies epidemic processes over discrete-time periodic time-varying networks. We focus on the susceptible-infected-susceptible (SIS)  model that  accounts for a (possibly) mutating virus. We say that an agent is in the disease-free state if it is not infected by the virus. %where the state of an agent can be interpreted as an approximation of the probability of the said agent being infected and say that the system is in the disease-free equilibrium (DFE) if every agent %\phil{is in the healthy state [this is kinda weird since we haven't defined `healthy state', i.e. defining one term with another term we haven't defined yet is not great]}.\\
 Our objective is to
devise a control strategy which ensures that all agents in a network exponentially (resp. asymptotically) converge to the disease-free equilibrium (DFE).  Towards this end, we first provide a) sufficient conditions for exponential (resp. asymptotic) convergence to the DFE; and b) a necessary and sufficient condition for asymptotic convergence to the DFE. The  sufficient condition  for global exponential stability (GES) (resp. global asymptotic stability (GAS)) of the DFE is in terms of the joint spectral radius of a set of suitably-defined matrices, whereas the necessary and sufficient condition for GAS of the DFE   involves the spectral radius of an appropriately-defined product of matrices. 
%over an \phil{interval of size equals length [I still can't understand this phrase, maybe needs a `the' or two, and/or `that is equal to the length'?]} of period. 
Subsequently, we leverage the 
stability results in order to design a distributed
% \said conditions in the design of a 
control strategy for eradicating the epidemic.% \phil{that addresses the aforementioned aims [I'm not sure this last phrase is  needed; it may be leading the reader a little too much, i.e., it's kinda obvious]}.}

% This is accomplished by first
% finding   conditions for exponential (resp. asymptotic) convergence to the DFE, independent of the initial state of each agent, that is, global exponential stability (GES) (resp. GAS). We provide,  in terms of the joint spectral radius of a set of matrices, a sufficient condition for  GES 
% % (resp. GAS) 
% of  the DFE, and, in terms of the spectral radius of a suitably-defined product of matrices, a necessary and sufficient condition  for global asymptotic stability (GAS) of  the DFE. 
% Based on these results, we present 
% a
% distributed  mitigation 
% strategy
% % strategies, 
% that 
% guarantees
% % guarantee 
% exponential (resp. asymptotic) convergence to the DFE. 
% Lastly,  simulations illustrate our theoretical findings and shed light on  the system behavior in the presence of noise,   
% violation of the necessary condition for GAS of the DFE, and  (possibly) improved distributed mitigation strategy. 

\end{abstract}

\section{Introduction}
Spreading processes, like epidemics,  propagation of (mis)information in social networks, etc., often have significant consequences. For instance, the  outbreak of  Severe Acute Respiratory Syndrome (SARS) in 2003 in Hong Kong resulted in $286$ deaths \cite{hung2003sars}. More recently, the increasing instances of coronavirus infections   have severely affected normal life across multiple continents \cite{dong2020interactive}. % with more than \seb{$6,400,000$} people being infected leading to around \seb{$380,000$} deaths  
%he number of deaths surpassing that of SARS \cite{munster2020novel}.
It is  known that certain epidemics exhibit yearly seasonal patterns, such as meningococcal meningitis in Western Africa, which typically occurs between January and May of each year \cite{pascual2005seasonal}. Furthermore, in the modern world, the networks that people have often recur with some periodicity, for example, professional networks during the day; personal networks at other times, and transportation networks.    In the present paper, we will consider seasonal epidemic processes in periodic time-varying networks, and will be interested in the following natural question: how can the epidemic be eradicated?
%or, at the very least, mitigated?
Answering this question is a two-step process: First, we need to know under what conditions do all the agents in a population become healthy. Second, given the knowledge of the convergence conditions, what measures can be adopted for guaranteeing that the epidemic gets eradicated.% (resp. mitigated).
%\phil{[don't we always eradicate in this paper? That is, do we ever mitigate the virus without completely removing it?]}}%Answering this question in the present day is all the more challenging since people have their own diverse networks --colleagues at work; friends and family at other times.}

%Moreover, certain epidemics exhibit seasonal pattern every year, for instance  meningococcal meningitis \cite{pascual2005seasonal}.}

%Consequently, following in the footsteps of Bernoulli's seminal paper \cite{bernoulli1760essai}
Modelling and analysis  of spreading processes has attracted the attention of researchers across a wide spectrum ranging from mathematical epidemiology  \cite{bernoulli1760essai,hethcote2000mathematics} and physics \cite{van2009virus} to  the social sciences \cite{easley2010networks}. %\phil{[If we have the citations on epidemiology we should probably have some on physics and social sciences too; it currently looks like we forgot]}.
The primary objective behind these research efforts is to better understand \emph{how} various diseases can spread through a population, which could then inform effective methods of management and control of the disease. 
In this pursuit, various models have been studied in the literature; %susceptible-infected-recovered (SIR), susceptible-exposed-infected-recovered (SEIR) and susceptible-infected-susceptible (SIS) models -- the first of which was developed  in \cite{kermack1932contributions}.\\
here, we concern ourselves  with susceptible-infected-susceptible (SIS) models.

In an SIS model, an agent is either in the susceptible or infected state.
% ; if the agent is susceptible, or healthy, it might become infected depending on whether or not it is exposed to the disease,  or it is in the infected state. 
A healthy agent can, as a consequence of its neighbors being infected, become infected with some infection rate $\beta$. An infected agent can be cured, with a healing rate $\delta$, thereby returning  to the susceptible state. It is assumed that there is no entry into or
% , with the (possible) exception of death, 
exit from the population, 
that is, the number of agents in the network remains fixed \cite{ahn2013global,pare2018epidemic}.

%\seb{In the present paper, since we concern ourselves with seasonal epidemics,  our focus will be on periodic SIS models. Devising strategies for eradicating (resp. mitigating) the spread of an epidemic typically involves the following steps: Find conditions such that all agents, regardless of their initial state, converge to the healthy )}

\subsection*{Related Works}
The analysis of SIS epidemic models has attracted the attention of researchers over the last several decades; for the   continuous-time case, see %for instance 
\cite{khanafer2016stability,fall2007epidemiological,van2009virus}, whereas for the  discrete-time case, see %for instance 
\cite{ahn2013global}, \cite{wang2003epidemic,peng2010epidemic,pare2018analysis}. In the present  paper, we consider a discrete-time setup, and therefore mainly review the discrete time literature. In this context, for the case of time-invariant graphs, \cite{wang2003epidemic} provides an epidemic threshold for the model equal to the inverse of the maximum eigenvalue of the matrix representing the graph structure. %The authors show that if the threshold is less than the ratio of the infection and healing rates then the virus will die out. 
However, the result in \cite{wang2003epidemic} is restricted to homogeneous virus spread, i.e., the infection and healing rates of each agent are identical.   \seb{The result in \cite{wang2003epidemic} has been further strengthened by  accounting for directed and weighted graphs  in \cite{peng2010epidemic}.  In particular, \cite[Theorem~5]{peng2010epidemic}
establishes that as long as the spectral radius of an appropriately-defined matrix is strictly less than one, the epidemic becomes extinct. However, note that the time-invariant discrete-time SIS model in \cite{peng2010epidemic} is different from the %time-invariant discrete-time SIS model 
one in \cite{pare2018analysis}.}
%provides a sufficient condition for   exponential stability of the healthy state, in terms of the spectral radius of the corresponding parametrized adjacency matrix.}
%within the scope of  time-invariant graphs and for a particular case of the homogeneous (i.e., the infection and healing rates of each agent are identical) virus model, \cite{wang2003epidemic} provides an epidemic threshold for the model equal to the inverse of the maximum eigenvalue of the matrix representing the graph structure. The authors show that if the threshold is less than the ratio of the infection and healing rates then the virus will die out.
% in terms of the maximum eigenvalue -- proportional \phil{[what is proportional? The maximum eigenvalue? The threshold? The main result in that paper is the threshold $\tau = 1/\lambda(A) < \beta/\delta$, this is not clear from this sentence]} to the ratio of the infection and healing rates -- of the matrix representing the graph structure.  
%The result in \cite{wang2003epidemic} has been further strengthened by  accounting for directed and weighted graphs  in \cite{peng2010epidemic}.  
%Peng \emph{et al.,} in \cite{peng2010epidemic}, strengthen the results in \cite{wang2003epidemic} by also admitting directed and weighted graphs.
The disease-free equilibrium (DFE) and the non-disease free equilibrium (NDFE)\footnote{The NDFE is an equilibrium where the infection persists in the network, and is also referred to as the \emph{endemic} equilibrium elsewhere in the literature; see for instance 
\cite{liu2019analysis,pare2018analysis}.} 
% -- that is, an equilibrium where the infection persists in the network -- 
of several models have been studied in \cite{ahn2013global}. Moreover, \cite{ahn2013global} also provides existence, stability and uniqueness conditions for the NDFE.
%Ahn and Hassibi, in \cite{ahn2013global}, study the disease-free equilibrium (DFE) and the non-disease free equilibrium (NDFE)\footnote{The NDFE is also referred to as the \emph{endemic} equilibrium elsewhere in the literature; see for instance \cite{liu2019analysis,pare2018analysis}.} -- that is, an equilibrium where the infection persists in the network -- of several models, and provide existence, stability and uniqueness conditions for the NDFE.
A necessary and sufficient condition,  in terms of the spectral radius of a matrix that is a function of the graph structure and the infection and healing rates, %suitably-defined matrix,  
for global asymptotic stability (GAS) of the DFE has been established in  \cite{pare2018analysis}. %}
%A more general case is the   one where \emph{multiple} competing viruses could be simultaneously active within a population;  see for instance  \cite{liu2019analysis,santos2015bi}. For the bi-virus case, the continuous-time setting has been addressed in \cite{liu2019analysis}, although the conditions therein guarantee only  \emph{asymptotic} convergence.
 To the best of our knowledge, for discrete-time time-invariant SIS models \seb{as in \cite{pare2018analysis}}, a sufficient condition for global exponential stability (GES) of the DFE is missing in the existing literature.
% \\

%\subsection*{Time-Varying Graphs}
 The models in \cite{ahn2013global}, \cite{wang2003epidemic,peng2010epidemic,pare2018analysis} %-- although contributing significantly towards our understanding of virus spread, and, consequently, towards development of  suitable eradication/mitigation strategies  -- 
 suffer from the following limitation: they cannot account for highly complex settings, in particular one where the interconnection between agents in a population (possibly) changes with time, for instance real-world social and human-interaction networks. Such a scenario imposes  a time-varying topology on the underlying graph, thus motivating the need for \emph{time-varying} SIS models. %\seb{That is, SIS models where the number of agents remains fixed, but the interconnection between them (possibly) changes with time.}

 The interest in SIS models with time-varying topology is  rather recent; for continuous-time setting see \cite{rami2013stability,pare2018epidemic,acc_multi}, while for discrete-time setting see \cite{prakash2010virus,bokharaie2010spread}. %Prakash  \emph{et al.,} in \cite{prakash2010virus}, by suitably improving upon the model in \cite{wang2003epidemic},   study a discrete-time  model, and provide a sufficient condition for \emph{local} exponential stability of the DFE.\\
 In the context of switched SIS models (both continuous-time and discrete-time), a sufficient condition for   local exponential stability (resp. instability) of the DFE is provided   in \cite{bokharaie2010spread}.  It turns out that the  condition in \cite[Theorem~2.2]{bokharaie2010spread} implies GAS of the DFE for a continuous-time switched SIS model; see  \cite{rami2013stability}.
%  strengthened said result  by showing 
  Following up on the work in \cite{rami2013stability}, \cite{liu2016threshold} studies a switching susceptible-infected (SI) model, albeit under the assumption of complete connectivity. In a similar vein, for a subset of random graphs, sufficient conditions for almost sure GES of the DFE are provided    in \cite{ogura2016stability}.
% ; see Theorem~5.1, but the setting therein is of continuous-time switched systems}. 
%Par\'e \emph{et al.,} in  \cite{pare2018epidemic},
In the continuous-time setting, for  heterogeneous virus spread and directed graphs, under  assumptions that the  topology of the underlying graph does not change \emph{too quickly},  
%\phil{Theorem 1 doesn't have the slow time varying assumption, right?])}
 sufficient conditions for exponential convergence to the DFE are provided in \cite[Theorem~2]{pare2018epidemic}. The setup considered in the present paper differs from the aforementioned works in the following sense: First,  we consider periodic discrete-time time-varying SIS models. \seb{That is, SIS models where the number of agents remains fixed, but the interconnection between them (possibly) changes with time.} Second,  we account for \seb{mutating viruses, that is, even the healing (resp. infection) rate of each agent can change with time. Moreover, the interconnections between agents, and the healing (resp. infection) rates repeat after every period.} %of some arbitrary length.  
As a first step towards designing a control strategy for eradicating   epidemics in the aforementioned setup, we ask the following questions: a) what are the sufficient conditions for the DFE to be GES? b) what are  necessary  and sufficient conditions for the DFE to be GAS? To the best of our knowledge, both the stated questions remain open. The present paper aims to answer these questions. 
The second step essentially involves comprehending how the dynamics of the spreading process can be \emph{controlled} so as to ensure that all agents converge to the DFE exponentially (resp. asymptotically) fast.
%In order to design an effective control strategy for eradicating (resp. mitigating the spread) of epidemics, it is crucial to 
  %  another line of investigation in mathematical epidemiology concerns  \emph{controlling} the dynamics of the spreading processes.
 % For the   eradication (resp. mitigation of spread) of infectious diseases, \\
In this regard, various strategies have been proposed in the literature; see for instance  \cite{liu2019analysis,enns2012optimal}, whereas for a survey of this subtopic, see \cite{nowzari2016analysis}. \seb{In particular,  \cite{enyioha2015distributed} considers a directed, network
comprising heterogeneous agents, and %seeks to control the spread of a virus, Towards this end, the authors in \cite{enyioha2015distributed}
proposes  a fully distributed  Alternating Direction Method of Multipliers (ADMM) algorithm that allows for local computation of optimal investment required to boost the healing rate at each
node. However, the ADMM algorithm in \cite{enyioha2015distributed} involves heavy communication overhead, since every agent needs to share with its neighbors its local estimate of the full network. Overcoming the drawbacks with the ADMM algorithm in \cite{enyioha2015distributed},  \cite{ramirez2018distributed} proposes distributed discrete-time nonlinear algorithms to handle a class of distributed resource allocation problems. A  decentralized algorithm that involves disconnecting nodes and increasing the healing rate subject to resource constraints has been  proposed in \cite{torres2016sparse}. This algorithm also accounts for %the sparse control paradigm; 
control sparsity, that is, control resources can be allocated only to a subset of nodes, and not necessarily to the whole network. A distributed algorithm that, given resource limitations, ensures eradication of an epidemic with a \emph{specified} rate has been recently provided in \cite{mai2018distributed}.
The distributed control algorithms provided in \cite{liu2019analysis},\cite{enyioha2015distributed, ramirez2018distributed,torres2016sparse,mai2018distributed}, and the results in \cite{enns2012optimal,nowzari2016analysis} are for time-invariant SIS models.} 
    %\phil{[given that we have the title ``distributed control" we should cite \cite{mai2018distributed}]}
%   \phil{[should we explain the control technique in \cite{pare2018epidemic}?]}
%   \seb{Response: The Introduction is already a bit long. Hence, I would avoid it.}
  Similar techniques for the more general setting of the discrete-time, periodic, time-varying, mutating SIS model are, as far as we know,   not available in the %existing
  literature; the present paper closes this gap.
%   
%\subsection*{Importance of the problem from a practical standpoint}

\subsection*{Contributions}
%Recall that 
The central premise of the paper is: given that a seasonal epidemic is prevalent within a population with %(possibly)
time-varying interconnection between the agents, how do we eradicate   it? 
We answer this question in the following manner.
First, we find  
% the 
conditions which ensure that, regardless of the initial state of an agent, i.e., healthy or sick, \emph{all} agents should converge to the healthy state exponentially (resp. asymptotically) fast. 
%\phil{[I deleted `the' before `conditions']}
%exponential (resp. asymptotic) convergence to the healthy state.
Second, with the knowledge of the aforementioned conditions in hand, we show that by sufficiently boosting the healing rate of each agent the epidemic can be eradicated in exponential (resp. asymptotic) time. More specifically, \philnew{under assumptions of periodicity, we show that:} %we tackle these problems in the following manner:
\begin{enumerate}[label =(\roman*)]
    \item \label{cont:1} %We show that, \seb{under assumptions of periodicity}, \\
   %We show that
   The joint spectral radius of an appropriately-defined set of matrices being strictly less than one  ensures GES  of the DFE; see Theorem~\ref{result1:GES}.
    \item \label{cont:2} %We show that, \seb{under assumptions of periodicity},\\
   % We show that
    The joint spectral radius of an appropriately-defined set of matrices being  no greater than one ensures  GAS of the DFE; see  Theorem~\ref{result:alt:GAS}. %\\
    A less restrictive condition that endows the DFE with the GES (resp. GAS) property requires the spectral radius of a suitably-defined product of matrices to be strictly less than (resp. not greater than) one; see Corollary~\ref{result2:GES} (resp. Corollary~\ref{thm:suff:AS}).
        In particular, for discrete-time SIS \emph{time-invariant} models, we establish
    that the spectral radius of a suitably-defined \emph{matrix} being strictly less than one implies GES of the DFE; see Proposition~\ref{prop:TI:GES}.
        
        % Given that even for discrete-time \emph{time-invariant} SIS models condition(s) ensuring GES of the DFE have insofar not been provided in the existing literature, Theorem~\ref{result1:GES} and Corollary~\ref{result2:GES} are of particular importance. Specializing the condition in Corollary~\ref{result2:GES} for time-invariant graphs, 
  
    \item  \label{cont:3}%We show that, \seb{under assumptions of periodicity},\\
    %We show that 
    The spectral radius, of a suitably-defined  product of matrices,  being no greater than one is  a necessary and sufficient condition for GAS of the DFE; see Theorem~\ref{Charzn:DFE}.
    \item \label{cont:4}  %Finally, we provide distributed control strategies that %, \seb{under assumptions of periodicity},\\
    A novel distributed control strategy exponentially (resp. asymptotically) stabilizes the DFE; see Theorem~\ref{thm:local:control:ensures:GES} (resp. Corollary~\ref{cor:local:control:ensures:GAS}).
\end{enumerate}
\subsection*{Outline}
The  paper unfolds as follows: we conclude the present section by listing all the notation used in the sequel. The problems under investigation, and hence, the main objectives of the paper, are %precisely
stated in Section~\ref{sect:pbform}, whereas the %necessary
background material needed for developing the main results are  in Section~\ref{sect:prelims}. We present conditions for exponential convergence (resp. asymptotic convergence) to the DFE in Section~\ref{sec:exp} (resp. Section~\ref{sec:asm}). We propose the distributed control strategy in Section~\ref{sec:con}.
%The main contributions, namely Theorems~\ref{result1:GES}, \ref{result:alt:GAS}, and \ref{Charzn:DFE} -- all of which deal with the  various stability properties of the DFE -- are provided in Section~\ref{sect:main-result}.
The simulations are provided in Section~\ref{sect:simulations}. Finally, we summarize the paper, and highlight certain problems that could be of possible interest for future work in Section~\ref{sect:conclusion}. 
\subsection*{Notation}
Let $\mathbb{R}$ (resp. $\mathbb{Z}_{\geq 0}$) denotes the set of real numbers (resp. non-negative integers). \seb{We denote by $\mathbb{Z}_+$ the set of positive integers. For a pair of integers $a,b \in \mathbb{Z}_+$, $a \text{ mod }b$ indicates $a$ modulo $b$.} For any positive integer $n$, we have $[n] = \{1,\hdots, n\}$ and $[n]^- = \{0,\hdots,n-1\}$. %, or something similar? It could be useful...]}
% \seb{Response: Agreed. Will change accordingly.}
Given a matrix $A \in \mathbb{R}^{n \times n}$, $a_{ij}$ denotes the entry corresponding to the $i^{th}$ row and $j^{th}$ column; and $\rho(A)$ denotes its spectral radius. Given a matrix $A$, supposing its spectrum is real, $\lambda_{\min}(A)$ (resp. $\lambda_{\max}(A)$) denotes the minimum (resp. maximum) eigenvalue of $A$.  A diagonal matrix is denoted as $\diag(\cdot)$. G
iven a vector $x \in \mathbb{R}^{n}$, its transpose is denoted as $x^\top$ \phil{and its average as $\bar{x}:=\frac{1}{n}\sum_{i=1}^n x_i$}.  The Euclidean norm is denoted by $\left\|\cdot\right\|$, whereas the infinity norm is indicated by $\left\|\cdot\right\|_{\infty}$. Given a sequence of matrices $A(k+p)$, $A(k+p-1)$, $\hdots$, $A(k+1)$, $A(k)$, their product  $A_{k+p+1:k}$ is defined as $A_{k+p+1:k}= A(k+p)\cdot A(k+p-1)\cdots A(k+1)\cdot A(k)$. Given a matrix $A$, $A \prec 0$ (resp. $ A\preccurlyeq 0 $) indicates that $A$ is negative definite (resp. negative semidefinite), whereas  $A \succ 0$ (resp. $ A\succcurlyeq 0 $) indicates that $A$ is positive definite (resp. positive semidefinite).

\section{Problem Formulation}\label{sect:pbform}

Consider a (possibly) time-varying epidemic network of $n$ agents, where the interpretation of \emph{time-varying} is as follows: the set of agents remains fixed, whereas the interconnections among the agents could (possibly) be time-varying.  Due to the (possibly) time-varying  nature of the interconnections, the healing rate and infection rate of each agent might also be time-dependent, that is, mutating. Thus, the continuous-time dynamics of each agent can be represented as follows:
%Consider the following continuous-time model: 
\begin{equation} \label{eq:ct}
\dot{x}_{i}(t) = (1-x_{i}) \beta_{i}(t)\sum\limits_{j=1}^{n}a_{ij}(t)x_{j}- \delta_{i}(t)x_{i}(t),
\end{equation}
where $i$ represents the $i^{th}$ agent, $x_{i}$ is the infection level,  and for every $t \in \mathbb{R}$, $\beta_{i}(t) \seb{\geq} 0$ (resp. $\delta_{i}(t) \seb{ \geq }0$) denotes the infection (resp. healing) rate. 
%\phil{[So why not do it here?]}
\seb{Assuming there exists a directed edge from agent $j$ to agent $i$ at time $t$, the corresponding edge weight %between any two agents $i$ and $j$, at time $t$, 
is denoted by $a_{ij}(t) > 0$. If $a_{ij}(t) =0$, then there does not  exist an edge from agent $j$ to agent $i$ at time $t$.} %\phil{, and therefore the set of agents, $V = \{1,2, \hdots, n\}$, and the set of edges, $E_k = \{(x_{i}, x_{j})\mid \beta_{i}(k)a_{ij}(k) \neq 0\}$, define a spread graph $G_{k} = (V, E_{k})$. [or something like this]} 
%\seb{
Intuitively, one can think of $x_i$ as an approximation of the probability of agent $i$ being infected, and $1-x_i$ represents an approximation of the probability of agent $i$ being healthy. The state can also be interpreted as the proportion of subpopulation $i$ that is infected.
Therefore, for the remainder of the paper we assume that the initial values of each agent lie in the interval~$[0,1]$.
%}

\seb{The model in~\eqref{eq:ct} was %previously
introduced in \cite[Equation~10]{pare2018epidemic}, whereas,  
 %particularized
 for the time-invariant case, it has been also proposed  in \cite[Equation~1]{pare2018analysis}.}

%\seb{
The virus outbreaks that motivate this work are often recorded in epidemiological reports that are compiled 
% the spread data 
per day \cite{whoCoronavirus,snow1855mode} or week \cite{whoEbola}. 
This sampling of the system behavior motivates the use of a discrete-time SIS model \cite{pare2018analysis}.
The model is
obtained by applying Euler's method \cite{atkinson2008introduction} to \eqref{eq:ct}, %is the following:
%}
\begin{equation} \label{eq:dt}
\begin{split}
    x_{i}(k+1) = x_{i}(k) + h\bigg( (1-x_{i}(k)) \beta_{i}(k)\sum\limits_{j=1}^{n}a_{ij}(k)x_{j}(k) \\
    - \delta_{i}(k)x_{i}(k)\bigg),
\end{split}
\end{equation}
\vspace{-3ex}

\noindent
where $h$ is the sampling parameter, \seb{and, therefore, $h >0$}. %\\
% The discrete time model is of interest because disease spread data is naturally sampled . 
Observe that system~\eqref{eq:dt} is a  discrete-time nonlinear time-varying system, and quite naturally its stability analysis differs considerably from that of discrete-time linear time-varying systems.
% \\

%Notice that, particularised for the time-invariant case, the continuous-time dynamics (as in Eq.~\eqref{eq:ct}) is developed by using mean-field approximation of a Markov chain model; see \cite{pare2018analysis,van2009virus}.
%Since the discrete-time version of~\eqref{eq:ct} is obtained by applying Euler discretization, \eqref{eq:dt} is an approximation of an approximation \cite[Remark~1]{pare2018analysis}. The approximation accuracy for the difference equation~\eqref{eq:dt} has been addressed, via simulations, in \cite[Section~2.2.2]{pare2018virus}. \phil{[Did Kalle want us to change this?]} \seb{A discussion on the intersample behavior accuracy is beyond the scope of the present paper. }

%\phil{[why not have this earlier again?]}
%\seb{Response: Because we need $\beta_{i}(k)a_{ij}(k)$ while defining the edge set.}
%\phil{[See above]}
The spread of diseases in a network can be modeled using a graph: the nodes representing the agents, and the edges representing the interaction among them. More formally, let $G_{k} = (V, E_{k})$ represent such a  network, where $V = \{1,2, \hdots, n\}$ is the vertex set, and \seb{$E_k = \{(x_{i}, x_{j})\mid %\beta_{i}(k)
a_{ij}(k) \neq 0\}$ is the edge set}. 

The model in \eqref{eq:dt} can be written in a matrix form as:
\begin{equation} \label{eq:matrixform1}
x(k+1) = x(k) + h ((I-X(k))B(k)A(k)-D(k))x(k)
\end{equation}
where $X(k) = \diag(x(k))$, $B(k) = \diag(\beta_{i}(k))$,  $D(k) = \diag(\delta_{i}(k))$, and $A(k) = [a_{ij}(k)]$, for every $i, j \in [n]$. Let us define $\bar{B}(k) := B(k)A(k)$, with its entries being denoted as $\bar{\beta}_{ij}(k)$. Then \eqref{eq:matrixform1} can be rewritten as:
\begin{equation} \label{eq:matrixform}
x(k+1) = x(k) + h ((I-X(k))\bar{B}(k)-D(k))x(k).
\end{equation}
\seb{Observe that $\bar{\beta}_{ij}(k)$ represents the infection rate and nearest-neighbor graph both %structure together
at time instant $k$. That is, assuming $\beta_{i}(k) > 0$ and that there exists an edge from vertex~$j$ to vertex~$i$ (i.e., $a_{ij}(k) > 0$), $\bar{\beta}_{ij}(k)$ scales the weight on the edge from vertex $j$  to vertex $i$ by $\beta_{i}(k)$.}

Since we are interested in seasonal epidemics, we restrict our attention to discrete-time \emph{periodic} SIS models. Thus, we have the following assumption. %Towards this end, we narrow our attention to periodic systems.  %with periodicity $p$, where $p \in \mathbb{Z}_{+}$.

%\phil{[We should reference the motivating applications from the Introduction here again to justify why periodicity is important]}
% That is, $B(k+p) = B(k)$, $A(k+p) = A(k)$, and $D(k+p) = D(k)$ for all $k \geq 0$}

\begin{assm}\label{assm:1}
Given some \philnew{period} $p \in \mathbb{Z}_{+}$, $B(k+p) = B(k)$, $A(k+p) = A(k)$, and $D(k+p) = D(k)$ for all $k \geq 0$
\end{assm}

The DFE  is defined as the state where $x_{i}(k) = 0$ for all $i \in [n]$, which, from \eqref{eq:matrixform}, implies that $x_{i}(\kappa) =0$ for all $\kappa \geq k$, for all $i \in [n]$. We are interested in ensuring that, irrespective of the initial condition of an agent, i.e., healthy or sick, the system should exponentially (resp. asymptotically) converge to the DFE. %\\
% \seb{
Throughout this paper, 
% depending on the context 
we interchangeably use the terms \enquote{healthy state} and \enquote{DFE}, and likewise the terms \enquote{convergence to the DFE} and \enquote{eradication of the virus.}
% }\\

With the  above-described setup in place, the objectives of the present paper are as follows:
\begin{enumerate}[label=(\roman*)]
\item \label{q1} For the system with dynamics as given in \eqref{eq:matrixform}, find sufficient condition(s) such that the DFE  is the only equilibrium and GES;

\item \label{q2} For the system with dynamics as given in \eqref{eq:matrixform}, find necessary  and sufficient condition such that the DFE  is the only equilibrium and GAS;

\item Based on the knowledge of the graph topologies, infection  rates and  the conditions for exponential (resp. asymptotic) convergence to the DFE, develop a distributed control strategy such that the DFE can be exponentially (resp. asymptotically) stabilized.
\end{enumerate}
% \\

\phil{We make the following assumptions.}

%The following assumptions are required for our model to be well-defined:
% \begin{assm}\label{assm:1}
% For all $i \in [n]$, we have $x_{i}(0) \in [0,1]$.~$\blacksquare$
% \end{assm}
% Intuitively, one can think of $x_i$ as an approximation of the probability of agent $i$ being infected, and $1-x_i$ represents an approximation of the probability of agent $i$ being healthy. Therefore, one can quite naturally assume that the initial values of each agent would lie in the interval $[0,1]$.
\begin{assm}\label{assm:2}
We have  $h\delta_{i}(k) \geq 0$ and $\bar{\beta}_{ij}(k) \geq 0$ for every $i, j \in [n]$, $k \in [p]^-$.~$\blacksquare$ 
\end{assm}

\begin{assm}\label{assm:3}
For every $i, j \in [n]$ and $k \in [p]^-$, $h\delta_{i}(k) \leq 1$ and $h\sum_{j}\bar{\beta}_{ij}(k) \leq~1$.~$\blacksquare$ 
% For every $i, j \in [n]$, $h\delta_{i}(k) \leq 1$ and $h\sum\limits_{j}\bar{\beta}_{ij}(k) \leq~1$, where $k \in [p]^-$.~$\blacksquare$ 
\end{assm}
Assumption~\ref{assm:2} says that, for each agent, the healing and infection rates are nonnegative. Assumption~\ref{assm:3} is required for ensuring that our model  is  well-defined. %approximation of~\eqref{eq:ct}}.}% as will be seen in the sequel.} \phil{[Since we don't include the proof of Lemma 1 the necessity of Assumption 3 is not clearly shown in the sequel]}
% TBD: Comment on Assumption~\ref{assm:3}.}
\begin{lem} \label{lem:eqm} For the system in~\eqref{eq:matrixform}, under the conditions of 
Assumptions~2 and~3 and if $x_{i}(0) \in [0,1]$, for all $i \in [n]$, then
$x_{i}(k) \in [0,1]$ for all $i \in [n]$ and $k \geq 0$.~$\blacksquare$
\end{lem}
% \textit{Proof:} 
The proof is along similar lines as that of  \cite[Lemma~1]{pare2018analysis}, and, hence, is skipped.~$\square$\\
Lemma~\ref{lem:eqm} ensures that the set $[0,1]^n$ is positively invariant, i.e., once a trajectory of~\eqref{eq:matrixform} enters the set $[0,1]^n$, it stays within the set $[0,1]^n$ for all future time instants. 

\begin{comment}
% \noindent 
With respect to the spread of virus, so as to rule out trivial  cases, we make the following assumption.
%\noindent The following assumption ensures nontrivial virus spread. 
\begin{assm}\label{assm:4}
We have $h \neq 0$ and, for all $k \in [p]^-$, there exists $i \neq j$ such that $\bar{\beta}_{ij}(k) > 0$.~$\blacksquare$
\end{assm}

\end{comment}

%\phil{[what if the graph was completely disconnected for one $k \in [p]^-$? It would still be nontrivial but violates this assumption (this would cause difficulties with the irreducibility assumption but we could fix that)]}
% \seb{Response: fixed; moved to asymptotic convergence subsection}

\section{Preliminaries}\label{sect:prelims}
In this section, we recall various notions of stability of discrete-time deterministic systems \cite[Section~5.9]{vidyasagar2002nonlinear}, which will be used in the sequel. Additionally, we collect some useful results from the literature that facilitate the development of our main results.
% \\

Consider a system, described as follows:
\begin{equation}\label{eq:autosys}
x(k+1) = f(k, x(k)),
\end{equation}
where $f: \mathbb{Z}_{\geq 0} \times \mathbb{R}^{n} \rightarrow \mathbb{R}^{n}$ is  locally Lipschitz. We say that an equilibrium of~\eqref{eq:autosys}  is (uniformly) asymptotically stable if it is (uniformly) stable and (uniformly) attractive.
% \footnote{\seb{The equilibrium $x =0$ is said to be uniformly stable if, for each $\omega > 0$, there exists a $\tau = \tau(\omega)$ such that 
% $k_0 \geq 0$, $\lvert|x_0\rvert| < \tau(\omega)$ $\implies$ $\lvert|x_k\rvert| < \omega$ $\forall k \geq k_0$.}}
% \footnote{\seb{The equilibrium $x =0$ is said to be uniformly attractive if there exists $\zeta >0$ such that 
%  $\lvert|x_0\rvert| < \zeta$, $k_0 \geq 0$ $\implies$ $x_k \rightarrow 0$ as $k \rightarrow \infty$, uniformly in $k_0$, $x_0$.\\} \sebcancel{COMMENT: Does it not suffice to just point to the definitions in \cite[Sect.5.9, Page 265]{vidyasagar2002nonlinear}?}}
An equilibrium is said to be GAS (resp. globally uniformly asymptotically stable (GUAS))
% (GAS) 
if in addition to being asymptotically stable (resp. uniformly asymptotically stable) the system converges to that equilibrium for any initial condition. We recall a sufficient condition for  GUAS of an equilibrium of \eqref{eq:autosys}.
\begin{lem}{\cite[Section~5.9 Thm.~27]{vidyasagar2002nonlinear}}\label{prop:vidyasagar} The DFE of system~\eqref{eq:autosys} is GUAS if there is a  function  $V: \mathbb{Z}_{+} \times \mathbb{R}^{n} \rightarrow \mathbb{R}$ such that i) $V(k,0) =0$, and, for all $x \neq 0$, $V(k,x) > 0$, ii) $V$ is decrescent, and  radially unbounded, and iii) $-\Delta V$ (where the forward difference function $\Delta V: \mathbb{Z}_+ \times \mathbb{R}^n \rightarrow \mathbb{R}$ is defined as: $\Delta V(k,x) = V(k+1, x(k+1)) - V(k,x)$)  is positive definite.~$\blacksquare$
\end{lem}
A stronger notion of stability is that of GES, which is defined as follows:
\begin{defn}\label{defn:GES}
An equilibrium point of~\eqref{eq:autosys} is GES if there exist positive constants $\alpha$ and $\eta$, with $0 \leq \eta <1$, such that
\begin{equation}
\left\|x(k)\right\| \leq \alpha \left\|x(k_{0})\right\|\eta^{(k-k_{0})} \hspace{1mm} \forall k,k_{0} \geq 0, \forall x_{k_{0}} \in \mathbb{R}^{n}. \nonumber
\end{equation}
\end{defn}
%\phil{[our definition of GES is a little different because we're only concerned with the domain $[0,1]^n$ and not $ \forall x_{k_{0}} \in \mathbb{R}^{n}$...]}

We recall a sufficient condition for GES of an equilibrium of~\eqref{eq:autosys} in the following proposition:
\begin{lem}{\cite[Section~5.9 Theorem.~28]{vidyasagar2002nonlinear}}\label{thm:vidyasagar:GES}
Suppose there exists a function $V: \mathbb{Z}_{+} \times \mathbb{R}^{n} \rightarrow \mathbb{R}$, and constants $a,b,c >0$ and $p>1$ such that $a\left\|x\right\|^{p} \leq V(k,x) \leq b\left\|x\right\|^{p}$, $\Delta V(k,x) \leq -c\left\|x\right\|^{p}$, $\forall k \geq 0$, and $\forall x \in \mathbb{R}^{n}$, then $x=0$ is an exponentially stable equilibrium of \eqref{eq:autosys}.~$\blacksquare$
\end{lem}
The initial values are
% , recalling  from Assumption~\ref{assm:1}, 
in the domain $[0,1]^n$, since otherwise they do not correspond to reality for the model under consideration. Consequently, we can say that the DFE of system~\eqref{eq:matrixform} is  GES if the condition in Definition~\ref{defn:GES} (resp. Lemma~\ref{thm:vidyasagar:GES}) is satisfied for all $x_{k_{0}} \in [0,1]^n$. Similarly, we say that the DFE of system~\eqref{eq:matrixform} is GAS if the condition in Lemma~\ref{prop:vidyasagar} is satisfied for all $ x_{k_{0}} \in [0,1]^n$.

The following lemmas will be needed for proving the sufficiency 
results in the sequel. 
% More precisely, Lemmas~\ref{lem:rantzer} and \ref{lem:pare} are needed for sufficiency, while Lemma~\ref{lem:zhou} is useful for proving necessity.

\begin{lem}{\cite[Proposition~1]{rantzer2011distributed}}\label{lem:rantzer}
Suppose that $M$ is a nonnegative matrix such that $\rho(M)<1$. Then there exists a diagonal matrix $P \succ 0$ such that $M^\top PM-P \prec 0$.~$\blacksquare$
\end{lem}

\begin{lem}{\cite[Lemma~3]{pare2018analysis}}\label{lem:pare}
Suppose that $M$ is an irreducible  nonnegative matrix such that $\rho(M) =1$. Then there exists a diagonal matrix $P \succ 0$ such that $M^\top PM-P\preccurlyeq 0$.~$\blacksquare$
\end{lem}

% \seb{
The following proposition is used for proving the necessity result in the sequel.
\begin{prop}{\cite[Section~5.9 Theorem~42]{vidyasagar2002nonlinear}}\label{thm:vidyasagar:unstable}
Consider the autonomous system 
\begin{equation} \label{eq:atuonomous:time:invariant}
x(k+1) = f(x(k)). 
\end{equation}
Define $A = [\frac{\partial f}{\partial x}]_{x =0}$.  If $A$ has at least one eigenvalue with magnitude greater than one, then $x=0$ is an unstable equilibrium of~\eqref{eq:atuonomous:time:invariant}.~$\blacksquare$
\end{prop}

Given that the present paper concerns periodic systems, we now recall a result concerning the time invariance of the spectrum of the state transition matrix.
\begin{prop}{\cite[Page 157]{bittanti1986deterministic}} \label{prop:Bittanti}
Consider the discrete-time $p$-periodic time-varying autonomous system 
\begin{equation} \label{eq:atuonomous:LTV}
x(k+1) = A(k)x(k). 
\end{equation}
\end{prop}
Let $A_{k+p:k}$ denote the corresponding state transition matrix. The spectrum of  $A_{k+p:k}$ is independent of $k$.~$\blacksquare$

% }

% \begin{lem}{\cite[Lemma~3]{zhou2017asymptotic}}\label{lem:zhou} The system~\eqref{eq:autosys} is asymptotically stable if and only if $\lim\limits_{ k\rightarrow \infty}\left\|\Phi(k,k_{0})\right\| =0$.~$\blacksquare$
% \end{lem}

One of the  approaches towards studying stability issues in time-varying networks relies on the notion of \emph{joint spectral radius} -- first introduced by Rota and Strang in \cite{rota1960note} --  of a set of matrices; see for instance \cite{bokharaie2010spread,rami2013stability}.   In the sequel, we will explore the relation between the joint spectral radius of an appropriately-defined set of matrices and GES (resp. GAS) of the DFE.\\
\noindent We define the following:
\begin{align}
M(k) &:= I-hD(k)+h\bar{B}(k)\label{eq:M}\\
\hat{M}(k) &:= I + h ((I-X(k))\bar{B}(k)-D(k))\label{eq:Mhat}\\
M_{k+p:k} &:= M(k+p-1)M(k+p-2)\cdots M(k).\nonumber
\end{align}
Observe that linearizing system~\eqref{eq:matrixform} around the DFE yields a linear time-varying periodic system, whose state matrix is $M(k)$.\\ 
Let $\mathcal{M} = \{M(0), M(1), \hdots, M(p-1)\}$ denote a set of $p$ matrices $M(k)$, where $k \in [p]^-$. As was defined in \cite{bokharaie2010spread}, the joint spectral radius of $\mathcal{M}$, denoted by $\rho(\mathcal{M})$, is:
\begin{align} \label{jsr}
    \rho(\mathcal{M}) = \lim\limits_{ p\rightarrow \infty} \sup \norm[\big]{M(p-1)M(p-2)\cdots M(0)}^{\frac{1}{p}}, \\ \text{where } M(k) \in \mathcal{M}, \forall k \in [p]^-. \nonumber
    \end{align}
    % where  $R(k) \in \mathcal{R}$, for all $k \in [p]^-$.
\seb{That is,} $\rho(\mathcal{M})$  is the largest eigenvalue of the product of $p$ matrices in $\mathcal{M}$ amongst \emph{all} products of $p$ matrices in~$\mathcal{M}$.

% \section{Main Results}\label{sect:main-result}
% In this section, we first present a sufficient, but not necessary, condition for GES (resp. GAS) of the DFE -- this condition is relevant, since it strengthens the existing results in the literature, as we will see later in this section. Subsequently, we establish a  necessary and sufficient condition for GAS of the  DFE. Finally, we conclude by developing control strategies that guarantee exponential (resp. asymptotic) convergence to the DFE.

\section{Exponential Convergence to the DFE} \label{sec:exp}
In this section, we present  sufficient conditions for GES of the DFE. In the context of epidemic outbreaks, these conditions guarantee eradication of the epidemic exponentially fast.
Recalling the understanding of joint spectral radius in \eqref{jsr}, the following result gives a sufficient condition for  the DFE to be GES.
\begin{thm} \label{result1:GES}
Consider~\eqref{eq:matrixform} under Assumptions~\ref{assm:1}--\ref{assm:3}. If $\rho(\mathcal{M}) < 1$, then the DFE is GES.~$\blacksquare$
\end{thm}
\textit{Proof:} See Appendix, where additionally we %also
establish an upper bound on the rate of convergence.~$\square$\\
%  \seb{
% \\
% \phil{[I know it looks prettier like this but it may be more informative to have the rate in terms of the model parameters. And, if not, we should put the definitions of $\sigma_3$ and $\sigma_2$ into equation environments so we can reference them here]}
% \phil{this footnote was confusing because it looked like a square on the rate}
% \footnote{To see how this rate is obtained, refer to the proof of \cite[Theorem~23.3]{rugh1996linear}}. 
%\textit{Proof:} 
The result in Theorem~\ref{result1:GES}, albeit restricted to periodic systems, is relevant in its own right: Theorem~\ref{result1:GES} gives a sufficient condition for GES of the DFE, whereas  the same condition in \cite[Theorem~2.2]{bokharaie2010spread}, particularized for periodicity assumptions, guarantees only \emph{local} exponential stability of the DFE.  Moreover, the proof technique is entirely different.
%Second, our requirements are less stringent in the sense that we ask for $\rho(\mathcal{M}) \leq 1$, where $\rho(\mathcal{M})$ is defined as in \eqref{jsr} with $M(k) = I -hD(k) +h\bar{B}(k)$, while \cite{bokharaie2010spread} asks for $\rho(\hat{\mathcal{M}}) \leq 1$, where $\rho(\hat{\mathcal{M}})$ is defined as in \eqref{jsr} with $\hat{M}(k) = I + h((I-X(k))\bar{B}(k)-D(k))$.  Finally, 

% Theorem~\ref{result1:GES} is an improvement on \cite[Theorem~2.2]{bokharaie2010spread}, since the former ensures global  exponential stability of $x=0$, whereas the latter ensures only \emph{local} exponential stability of $x=0$. Moreover, the  requirements in  Theorem~\ref{result1:GES} are less stringent in the sense that it asks for $\rho(\mathcal{M}) \leq 1$, where $\rho(\mathcal{M})$ is defined as in \eqref{jsr} with $M(k) = I -hD(k) +h\bar{B}(k)$, while \cite[Theorem~2.2]{bokharaie2010spread} asks for $\rho(\hat{\mathcal{M}}) \leq 1$, where $\rho(\hat{\mathcal{M}})$ is defined as in \eqref{jsr} with $\hat{M}(k) = I + h((I-X(k))\bar{B}(k)-D(k))$.

Notice that checking the condition on the joint spectral radius in Theorem~\ref{result1:GES}  essentially entails asking the following question: given a set of matrices, say $\mathcal{R}$, is each and every product of matrices within $\mathcal{R}$ stable? Answering this is known to be NP-hard; see {\cite[Corollary~2]{tsitsiklis1997lyapunov}}. Hence, we are motivated to seek a different condition that is computationally tractable. %\\   

The following corollary is an  immediate consequence of the proof of Theorem~\ref{result1:GES} and the result in Proposition~\ref{prop:Bittanti}, and provides a less restrictive sufficient condition for GES of the DFE.

\begin{cor} \label{result2:GES}
Consider~\eqref{eq:matrixform} under Assumptions~\ref{assm:1}--\ref{assm:3}. If, for some $k \in [p]^-$, $\rho(M_{k+p:k}) < 1$, then the DFE is GES.~$\blacksquare$
\end{cor}

For the continuous-time setting,  \cite[Theorem~2]{pare2018epidemic} gives a sufficient condition for GES of  the DFE, under the assumption that the rate of change of topology is suitably bounded. %\phil{As mentioned in the Introduction,} \seb{in the current version, we dont mention this in the Introduction.}
To the best of our knowledge, for discrete-time time-varying SIS epidemics, Theorem~\ref{result1:GES} and Corollary~\ref{result2:GES} are first %-ever
%\phil{first-of-their-kind [may be a bit informal]}
results for GES of the DFE.
%In the context of  periodic time-varying setting,Theorem~\ref{result1:GES} and Corollary~\ref{result2:GES} may be viewed as discrete-time counterparts of \cite[Theorem~2]{pare2018epidemic}.
Moreover, unlike \cite[Theorem~2]{pare2018epidemic}, neither Theorem~\ref{result1:GES} nor Corollary~\ref{result2:GES} rely on any restrictions on \emph{how large} 
%\phil{[or how large? since the discrete time fixes the time change, amount of change may be analogous...]} 
the variations in topology can be.
% \\

%Given that the problem being investigated pertains to  mathematical epidemiology, it is of interest to
The following remark
provides an epidemiological interpretation of the implications of Theorem~\ref{result1:GES} (and Corollary~\ref{result2:GES}).
\begin{rem}\label{rem:math:epidem}
The result in Corollary~\ref{result2:GES} is useful in the following sense: %subject 
Subject to virus mutation and the underlying sequence of graph topologies repeating with some period $p$, we can conclude that the virus will be eradicated. Notice that this result is irrespective of how the aforementioned parameters vary with time; in particular, even if at times, compared to the healing rates, the infection rates are dominant. The same has been illustrated via simulations in Section \ref{sect:simulations}.~$\blacksquare$
\end{rem}
\seb{Recall that \cite[Theorem~5]{peng2010epidemic} provides a sufficient condition for  exponential convergence to the DFE. However, the model in \cite{peng2010epidemic} is different from~\eqref{eq:matrixform};  healing (resp. infection) rates are nonnegative and cannot be greater than one, and the edge-weights and the spreading parameters are not allowed to vary with time. 
Additionally, there are certain restrictions on the underlying time-invariant graph; in particular, all self-loops have weight equal to one, and no edge can have weight greater than one. Consequently, \cite[Theorem~5]{peng2010epidemic} is less general than Theorem \ref{thm:local:control:ensures:GES}.} %An implication of Corollary~\ref{result2:GES} is that for the discrete-time periodic time-varying SIS models the condition in \cite[Theorem~5]{peng2010epidemic} is not necessary. The same has been illustrated via simulations in Section \ref{sect:simulations}.}
% \begin{rem}[Epidemiological Interpretation]\label{rem:math:epidem}
% \seb{The result in Corollary~\ref{result2:GES} is useful in the following sense: 
% % a priori knowledge of neither the infection (resp. healing) rates nor the changes in topology of the underlying graph, are needed \phil{[I don't follow this sentence]}. Hence, 
% subject to virus mutation and the underlying sequence of graph topologies repeating with some period $p$, irrespective of how the aforementioned parameters vary with time -- in particular, even if at times, compared to the healing rates, the infection rates are dominant, as is illustrated via simulation in Section \ref{sect:simulations} --, we can conclude that the virus will be eradicated.}~$\blacksquare$%\\
% \end{rem}
%\phil{[I like this remark a lot, but it's one really long sentence. I think we need to break it up and make it a little clearer/easier to read. If you want me to do it let me know.]}

To the best of our knowledge, for \seb{the SIS model in~\eqref{eq:matrixform}, even when particularized for the \emph{time-invariant} case (thus yielding the model in \cite{pare2018analysis}),} %discrete-time \emph{time-invariant} SIS models,
a sufficient condition for exponential convergence to the DFE does not exist in the literature. It can be  immediately seen  that the condition in Corollary~\ref{result2:GES} can be specialized for the time-invariant setting, as discussed next. First note that time-invariant systems are periodic systems with periodicity \philnew{$p=1$}. Hence, $M_{k+p:k} =M$ for all $k$, which implies that the condition in Corollary~\ref{result2:GES} is satisfied if and only if $\rho(M) < 1$. Therefore we have the following proposition.

%\phil{[This corollary reads more like a proof. I think it should probably be rewritten more formally, especially if we make it a theorem, i.e. `Consider the non-mutating, static graph topology version of (4), that is, where p = 1. ...']}
\begin{prop}\label{prop:TI:GES}
Consider the non-mutating, static graph topology version of~\eqref{eq:matrixform}, that is, where $p = 1$. If $\rho(M)<1$, then the DFE is GES.~$\blacksquare$
\end{prop}
Proposition~\ref{prop:TI:GES} establishes GES of the DFE, whereas {\cite[Theorem~2]{pare2018analysis}},  under the same condition as in Proposition~\ref{prop:TI:GES}, establishes only GAS of the DFE. Hence, Proposition~\ref{prop:TI:GES} is a stronger version of {\cite[Theorem~2]{pare2018analysis}}.

%\seb{
Notice that, on one hand, the conditions in Theorem~\ref{result1:GES} and Corollary~\ref{result2:GES} involve strict inequalities. On the other, they guarantee faster convergence to the DFE. An obvious question that one can ask is the following: is it possible to relax the strict inequalities in Theorem~\ref{result1:GES} and Corollary~\ref{result2:GES} at \philnew{the cost of \emph{slower} convergence?} % to the DFE? 
In the context of epidemiology, the motivation for doing so goes along the following lines: depending on the severity of the epidemic in question, there might be scenarios, for instance the common cold, where a positive answer to the question: \enquote{will the disease die out?} suffices, and one is not too concerned with the \emph{speed} with which the epidemic disappears.  We investigate the same in the next section.%}
%\phil{[this is really good]}

%An obvious question that arises at this point is the following: what is the effect on the stability of the DFE if  the aforementioned inequalities were not necessarily strict? We investigate the same in the next section.

%\phil{[The transition here needs some work. We need to motivate why we're doing GAS after GES... I think we also need to talk a little more specifically about the motivating applications, i.e. ``are needed for asymptotically eradicating  the epidemic'' is probably too vague. This can probably be achieved by talking about what relaxing the strict inequality means in terms of the epidemiological interpretation, that is, why does it matter, in terms of the application, that we can do that. It will probably also help the transition if we end Section IV with a discussion of the importance of the Prop \ref{prop:TI:GES} in terms of virus spread application not just with relation to \cite{pare2018epidemic} (which could be a bit redundant after Remark 1 but I don't think Kalle will think so, if we write it well)]}

\section{Asymptotic  Convergence to the DFE}\label{sec:asm}
%\seb{
It turns out that if the inequality in Theorem~\ref{result1:GES} and Corollary~\ref{result2:GES} were to be not necessarily strict, then the DFE is GAS. Furthermore, if the inequality in Corollary~\ref{result2:GES} were to be reversed, then the healthy state is an unstable equilibrium. Thus,  in this section, we establish a sufficient condition, and  a necessary and sufficient condition for GAS of the  DFE.%} %These conditions are needed for asymptotically eradicating  the epidemic.

%
% Thus, in this subsection, we establish these claims, and show that the condition from Corollary~\ref{result2:GES} without strict inequality is both necessary and sufficient for GAS of the DFE.
% \\}
%\phil{[We just said that we have GES which is awesome and GAS is not as awesome, and now we're like, okay we can do GAS too, which seems a little strange. I'm not sure how to fix it but this transition should be a bit smoother (similar to the transition from Theorem 1 to Corollary 1). Maybe this subsection should be `Necessary and Sufficient Condtions to the DFE' or something... that may help explain the transition better]}

We begin  by noting that an immediate consequence  of  Theorem~\ref{result1:GES} is the following.
\begin{cor} \label{less:than:1:GAS}
Consider~\eqref{eq:matrixform} under Assumptions~\ref{assm:1}--\ref{assm:3}. If $\rho(\mathcal{M}) < 1$, then the DFE is GAS.~$\blacksquare$
\end{cor}

It turns out that   the DFE is endowed with the  property of  GAS even if  $\rho(\mathcal{M}) =1$. To prove this claim, we need, besides the assumptions in Corollary~\ref{less:than:1:GAS}, the following assumptions.
%With respect to the spread of virus, so as to rule out trivial  cases, we make the following assumption.
%\noindent The following assumption ensures nontrivial virus spread. 
\begin{assm}\label{assm:4}
We have $h \neq 0$ and, for all $k \in [p]^-$, there exists $i \neq j$ such that $\bar{\beta}_{ij}(k) > 0$.~$\blacksquare$
\end{assm}

 \begin{assm}\label{assm:5}
For each $k \in [p]^-$, the graph $G_k$ is strongly connected.~$\blacksquare$
\end{assm}
Assumption~\ref{assm:4} rules out scenarios wherein an agent is infected yet, since it is not connected to any of the other agents in the network, it does not transmit the virus. %, in which case there is \emph{no} virus spread.
 Assumption~\ref{assm:5} implies that the adjacency matrix $\bar{B}(k)$, where $k \in [p]^-$, is irreducible, i.e., $\bar{B}(k)$ cannot be permuted to a block upper triangular matrix.

%\phil{[This is only true if $\beta_{i}(k)>0$ for every $i \in [n]$ and $k \in [p]^-$]}
 \begin{prop} \label{equal:1:GAS}
Consider~\eqref{eq:matrixform} under \seb{Assumptions~\ref{assm:1}--\ref{assm:5}}. If $\rho(\mathcal{M}) = 1$, then the DFE is GAS.~$\blacksquare$
\end{prop}
\textit{Proof:} See Appendix~$\square$
% \\

Combining Corollary~\ref{less:than:1:GAS} and Proposition~\ref{equal:1:GAS}, we readily obtain the following result.
\begin{thm} \label{result:alt:GAS}
Consider~\eqref{eq:matrixform} under \seb{Assumptions~\ref{assm:1}--\ref{assm:5}}. If $\rho(\mathcal{M}) \leq 1$, then the DFE is GAS.~$\blacksquare$
\end{thm}

%\phil{[This is the exact same statement as Proposition 3. Is Proposition 3 only supposed to be equality?]}

\philnew{Theorem~\ref{result:alt:GAS} %also
establishes asymptotic stability of the equilibrium point $x =0$, %(i.e., the DFE) also for the case even for
including the case where $\rho(\mathcal{M}) =1$.} Thus, it differs from \cite[Theorem~2.2]{bokharaie2010spread} wherein no conclusions can be drawn when $\rho(\mathcal{M}) =1$.%\\

\sebcancel{It is well-known that the condition in Theorem~\ref{result:alt:GAS} is undecidable; see \cite[Theorem~2]{blondel2000boundedness}. That is, given a finite set of matrices $\mathcal{R}$, it is impossible to construct an algorithm that gives a correct binary answer to the question: is $\rho(\mathcal{R}) \leq 1$?} %Hence, we provide a  condition, different from that in Theorem~\ref{result:alt:GAS},  for GAS of the DFE.

Observe that  from Proposition~\ref{prop:Bittanti} and the proof of Theorem~\ref{result:alt:GAS}, the following, less restrictive, sufficient condition %for GAS of the DFE 
is immediate.

 \begin{cor} \label{thm:suff:AS}
Consider~\eqref{eq:matrixform} under \seb{Assumptions~\ref{assm:1}--\ref{assm:5}}. If, for some $k \in [p]^-$, $\rho(M_{k+p:k}) \leq 1$, then the DFE is  GAS.~$\blacksquare$
\end{cor}

%Given that the condition in Corollary~\ref{thm:suff:AS} is less restrictive,\\
\philnew{It turns out that condition in Corollary~\ref{thm:suff:AS} is also necessary.} 
%is worth investigating whether the condition in Corollary~\ref{thm:suff:AS} is necessary. % as well.
%The following proposition addresses the same.}
\begin{prop}\label{prop:nece:global:AS}
Consider~\eqref{eq:matrixform} under \seb{Assumptions~\ref{assm:1}--\ref{assm:5}}.
% Suppose that the system~\eqref{eq:matrixform} is $p$-periodic. 
The DFE is asymptotically stable only if,  for some $k \in [p]^-$,  $\rho(M_{k+p:k}) \leq 1$.~$\blacksquare$
\end{prop}
\textit{Proof:} See Appendix.~$\square$\\

Combining Corollary~\ref{thm:suff:AS} and Proposition~\ref{prop:nece:global:AS}, readily yields the following:
\begin{thm} \label{Charzn:DFE}
Consider~\eqref{eq:matrixform} under \seb{Assumptions~\ref{assm:1}--\ref{assm:5}}. The DFE of system~\eqref{eq:matrixform} is GAS if, and only if, for some $k \in [p]^-$, $\rho(M_{k+p:k}) \leq 1$.~$\blacksquare$
\end{thm}

For \philnew{the} continuous-time setting, it has been shown that the switched SIS model admits a limit cycle if the condition in Theorem~\ref{Charzn:DFE} is violated; see {\cite[Theorem~6.4]{rami2013stability}}. While our simulations suggest that,  for the discrete-time setting, violating the condition in Theorem~\ref{Charzn:DFE} could lead to the existence of a limit cycle (see Section~\ref{sect:simulations}),  this conjecture remains open. The main difficulty in proving this claim stems from the fact that the celebrated Poincar\'e-Benedickson Theorem -- which  forms the underpinning for the proof in the continuous-time case -- does not seem to have a discrete-time counterpart.

Indeed, Theorem~\ref{Charzn:DFE}, particularized for the case of time-invariant SIS models, coincides with  {\cite[Theorem~2]{pare2018analysis}}. To see this, consider the following:
\begin{rem}\label{rem:pare18:Lticase}
If system~\eqref{eq:matrixform} is time-invariant. or equivalently, $p=1$, then the condition in Theorem~\ref{Charzn:DFE} coincides with the condition in {\cite[Theorem~2]{pare2018analysis}}. To see this, consider the following argument: since $p=1$, for every $k \in \mathbb{Z}_{\geq0}$ $M_{k+p:k} =M$.  Therefore $\rho(M_{k+p:k}) = \rho(M)$. Hence $\rho(M_{k+p:k}) \leq 1$ if and only if $\rho(M) \leq 1$, which is the same as the condition in {\cite[Theorem~2]{pare2018analysis}}.~$\blacksquare$
\end{rem}
Notice that the objective insofar has been to find conditions that ensure exponential (resp. asymptotic) convergence to the DFE. Knowledge of the stability  conditions enables health administration officials to determine \emph{how} the model parameters in \eqref{eq:matrixform} should be adjusted so as to completely stop  the spreading of a disease. We focus on the same in the next section.  
%\phil{[Similar to my comment at the end of Section IV, we need a transition here that summarizes the results of this section in terms of the application and motivates the next section]}

\begin{comment}
\subsection{Mitigation Strategies}

%\phil{
% We now propose two different types virus mitigation algorithms that appeal to the analysis from Sections \ref{subsec:exp}-\ref{subsec:asm}. 
First, we propose a distributed algorithm that locally updates the healing rates as a function of susceptibility and nearest neighbor connections. 
Second, we propose a quarantine technique.

\end{comment}

\section{Distributed Control Strategy} \label{sec:con}
In this section, we show that increasing the healing rate of each agent by a sufficiently high amount ensures eradication  of the epidemic. For time-invariant continuous-time SIS models, (local)  techniques for eliminating the spread of epidemics have been provided in \cite[Section~\rom{5}]{liu2019analysis}. Inspired by the same, we explore similar strategies for the periodic mutating setup, as in the present paper. More specifically, in the sequel we study how to influence the healing rate of each agent so that the DFE is exponentially (resp. asymptotically) stabilized. Towards this end, we consider the following \philnew{healing rates}:
% }
 \begin{equation} \label{eq:local;control}
 \delta_{i}(k) = \sum\limits_{j=1}^{n}\bar{\beta}_{ij}(k) + \gamma_{i}, 
%  \text{ for } 
\forall
 k \in [p]^-,   i \in [n],
  \end{equation}
   where, for each $i \in [n]$, $\gamma_{i} >0$. %\\
   
It turns out that choosing healing rates  as in \eqref{eq:local;control} \seb{together with appropriate assumptions on $\gamma_i$} ensures that the DFE of \eqref{eq:matrixform} is GES, as evidenced by the following theorem.
\seb{\begin{thm} \label{thm:local:control:ensures:GES}
Consider~\eqref{eq:matrixform} under Assumptions~\ref{assm:1}--\ref{assm:2}. Suppose that, for each $i \in [n]$, $\gamma_i$ in~\eqref{eq:local;control} satisfies $h\sum\limits_{j=1}^{n}\bar{\beta}_{ij}(k) + h\gamma_{i} \leq 1$. Then  for healing rates as in~\eqref{eq:local;control}, the DFE of~\eqref{eq:matrixform} is GES.~$\blacksquare$
\end{thm}}
%\phil{Do we need Assumption 5? I think assuming just 1-4 is sufficient for the GAS case... The corollary will need Assumption 5.}
\textit{Proof:} See Appendix.~$\square$% where additionally we also establish an upper bound on the rate of convergence.
%  \seb{
 %The rate of convergence to the DFE for the control strategy given in \eqref{eq:local;control} may be obtained from Proposition~\ref{rem:decay:rate} in the Appendix. 
%  }
%\phil{[It seems a little strange to me to reference a Proposition that is in the Appendix. I understand that it depends on notation that is introduced in the Appendix but it's still a little weird...If we figured out a way to include it here, it would add more length to this section as well (which is not exactly necessary but wouldn't hurt]}
%\phil{COMMENT: We cannot appeal directly to the condition in Theorem~\ref{result1:GES}, since it deals with the joint spectral radius of a set of matrices, whereas with the control (16) we can only bound from above the spectral radius of $\tilde{M}^1$. However, from the proof of Theorem~\ref{result1:GES}, we can see--look at the paragraph starting with \enquote{Since, by assumptions 2-4,... marked with red}-- that 
 %$\rho(\tilde{M}^1) < 1$ implies that the DFE is GES.}
 
% It turns out that the DFE is (asymptotically) stabilizable even
\seb{ Observe that if $\gamma_{i} = 0$ $\forall i \in [n]$   in~\eqref{eq:local;control}, then, from the proof of Theorem~\ref{thm:local:control:ensures:GES}, 
% \phil{[Why not write `Proposition' out all the way? We have plenty of space.]} 
% it is immediate  that $\rho(\tilde{M}^1) \leq 1$. Subsequently, from 
and that of Proposition~\ref{equal:1:GAS}, it follows that the DFE is GAS, thus leading us to the following corollary.}
\begin{cor} \label{cor:local:control:ensures:GAS}
Consider~\eqref{eq:matrixform} \seb{under Assumptions~\ref{assm:1}--\ref{assm:5}}. For healing rates of the form
% ~\eqref{eq:local;control} with $\epsilon$ set to zero, 
\begin{equation*} %\label{eq:local;control2}
 \delta_{i}(k) = \sum\limits_{j=1}^{n}\bar{\beta}_{ij}(k), %\text{ for } k \in [p]^-,
\end{equation*}
for each $i \in [n]$ and $k \in [p]^-$, the DFE of~\eqref{eq:matrixform} is GAS.~$\blacksquare$
\end{cor}

The key insight that can be gleaned from Theorem~\ref{thm:local:control:ensures:GES} and Corollary \ref{cor:local:control:ensures:GAS} is that there exist sufficiently large, yet finite, time-varying healing rates such that the DFE can be stabilized. This could inform healthcare professionals of  disease-response techniques, as explained in the following remark. 
%  \seb{
\begin{rem} \label{rem:antidote} The distributed control strategy proposed in Theorem~\ref{thm:local:control:ensures:GES} (resp. Corollary~\ref{cor:local:control:ensures:GAS}) may be interpreted in the following sense: if the healing rate of each agent is suitably increased -- for instance by injecting  sufficiently high dosages of antidote %\phil{[I think it makes sense for this to be singular because one virus should probably have one antidote]}
--  then the virus will be eradicated exponentially (resp. asymptotically) fast. \seb{While the control strategy is extreme, i.e., no constraints are imposed, it informs decision makers of the best way to respond if ample resources are available and encourages the stockpiling of resources to combat possible outbreaks.}~$\blacksquare$
\end{rem}
\begin{figure}
  \centering
    \includegraphics[trim=400 125 200 125,clip,width = \columnwidth] {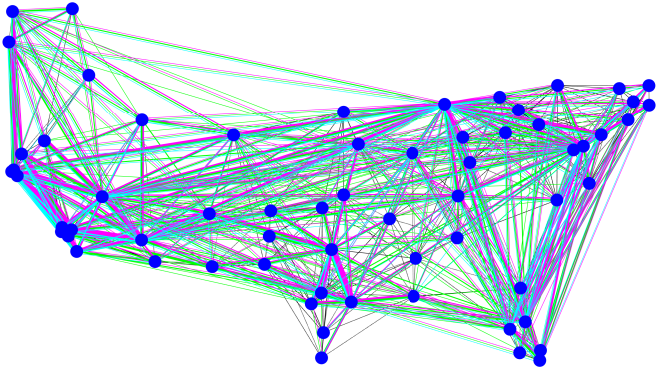}
   \caption{Final condition for simulation with $\delta = 35$. All nodes are in the DFE, depicted by blue.}
    \label{fig:healthy}
\end{figure}

\section{Simulations}\label{sect:simulations}

\seb{
The main challenge with new infectious \philnew{diseases} is that it is \philnew{often} unclear how they spread and how contagious they are \cite{hung2003sars}. Therefore, motivated by diseases like SARS and COVID-19 and
}
to further understand the implications of the results from the previous sections, % on epidemic processes,
we present %various
simulations over a network of 64 cities in the United States. 
The default graph structure is  a binary, nearest-neighbor graph, depicted in Figure \ref{fig:healthy} by black dotted edges. For the periodic parts of the network we aggregate
the Southwest Airlines flights between the cities, split by departure time in the morning (0:00-8:00), the day (8:00-16:00), and the \philnew{evening} (16:00-24:00), which are depicted in Figure \ref{fig:healthy} by green, magenta, and cyan, respectively and the edge weights are scaled by the number of flights.
\seb{The maximum number flights in and out of one city during the day is 125; therefore, we set $h=0.005$ to meet Assumption \ref{assm:3}.} 
We use the initial condition of Albuquerque completely infected and every other city completely healthy (however the results are independent of initial condition). 
% In this section we provide a set of simulations that illustrate the main results and some unproven behavior. 
In the plots of the network, blue ($b$) represents healthy and red ($r$) represents infected. The coloring of each node $i$ at time $k$ follows 
\begin{equation}
    x_i r + (1-x_i)b. \label{eq:color}
\end{equation}
% For the first two simulations we use homogeneous virus spread. 

% \subsection{Illustrative Simulations}

\begin{figure}
  \centering
    \begin{overpic}[trim=10 0 10 12,clip,width = \columnwidth] {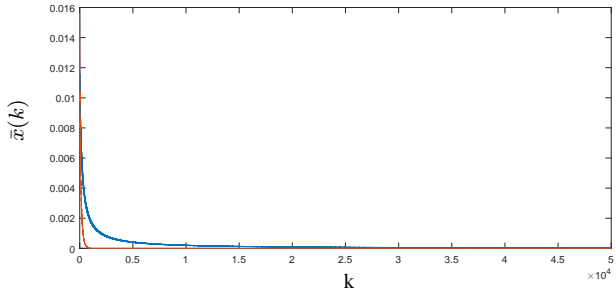}
            \put(51.5,-1){{\parbox{\linewidth}{%
                \footnotesize 
                k}}}
            \put(1,23.25){{\parbox{\linewidth}{%
                    \rotatebox{90}{\footnotesize $\bar{x}(k)$}}}}
        \end{overpic}%\includegraphics[trim=10 0 10 12,clip,width = \columnwidth] {usa_avg.eps}
   \caption{Average infection level of the cities over time. Blue is for the system with $\rho(M_{k+3:k}) = 1.000$ and red is for $\rho(M_{k+3:k}) < 1.000$.}
    \label{fig:avg}
\end{figure}

Since our interest lies in eradicating the virus, we explore the results on the stability of the DFE from Sections \ref{sec:exp} and \ref{sec:asm} via simulations. For simplicity we use homogeneous, non-mutating virus parameters with $\beta=1$. \seb{We also suppose that the healing rate of each agent is the same, i.e., $\delta_i = \delta$ for each $i \in [n]$.} Recall from Proposition~\ref{prop:Bittanti}, that the spectrum of $M_{k+p:k}$ is independent of $k$. When $\delta = 35$, $\rho(M_{k+3:k}) = 0.9815$. %for all $k \in \{0,1,2\}$, \seb{\enquote{which, recall, is in line with the result in Proposition~\ref{prop:Bittanti}}..COMMENT: I think the sentence in quotes can be dropped}\\
Consistent with the results in Section \ref{sec:exp}, the system converges to the DFE; see Figure \ref{fig:healthy}. When $\delta =33.765$, $\rho(M_{k+3:k}) = 1.000$. % for all $k \in \{0,1,2\}$.\\
In line with the results in Section \ref{sec:asm}, the system with $\rho(M_{k+3:k}) = 1.000$ converges at a much slower rate than the other; see Figure~\ref{fig:avg}.
For both of these systems $\rho(M_0) < 1$, while $\rho(M_1)$ and $\rho(M_2)$ are greater than one. Thus we see that even if the infection rates dominate the healing rates for the majority of the time, the virus can still be eradicated. This insight offers hope for control algorithm design; having actuator capabilities for some portion of the period might be sufficient to eradicate a virus. 

% \begin{figure}
%   \centering
%     \includegraphics[trim=10 0 10 12,clip,width = \columnwidth] {usa_avg.eps}
%   \caption{Average infection level of the cities over time. Blue is for the system with $\rho(M_{k+3:k}) = 1.000$ and red is for $\rho(M_{k+3:k}) < 1.000$.}
%     \label{fig:avg}
% \end{figure}

% \subsubsection{Limit Cycle Behavior}

% \phil{
% [The transition from the previous paragraph to this next one isn't that smooth... any suggestions would be appreciated]}
% \seb{Response: Here is one way to do it.}
 
Next, we focus on illustrating the instability result %[technically it's not written as an instabiltiy result...]}
from Section~\ref{sec:asm}. Our simulations, consistent with the result in Proposition~\ref{prop:nece:global:AS}, exemplify that  when $\rho(M_{k+p:k}) > 1$, the DFE is an unstable equilibrium. Moreover,   although it still needs to be rigorously proven, our simulations show the existence of limit-cycle behavior, thus implying  \emph{persistence} of an outbreak when $\rho(M_{k+p:k}) > 1$.
 
%However, there are serious concerns about the persistence of an outbreak when $\rho(M_{k+p:k}) > 1$. While it still needs to be proven, simulations show the existence of limit-cycle behavior.

\begin{figure}
  \centering
    \begin{overpic}[trim=10 0 10 12,clip,width = \columnwidth] {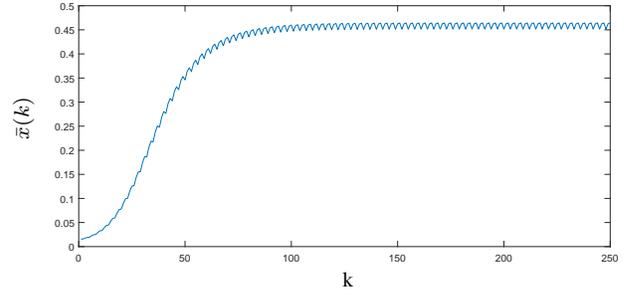}
            \put(51.5,-1){{\parbox{\linewidth}{%
                \footnotesize 
                k}}}
            \put(2,23.25){{\parbox{\linewidth}{%
                    \rotatebox{90}{\footnotesize $\bar{x}(k)$}}}}
        \end{overpic}%\includegraphics[trim=10 0 10 12,clip,width = \columnwidth] {usa_limit_avg.eps}
   \caption{Average infection level of the cities over time  for the simulation with $\delta_i(k) = 10$,  for every  $k \in \{0,1,2\}$, $ i \in [n]$. }
    \label{fig:avg_infected}
\end{figure}

\begin{figure}
   \centering
   \includegraphics[trim=400 125 200 125,clip,width = \columnwidth] {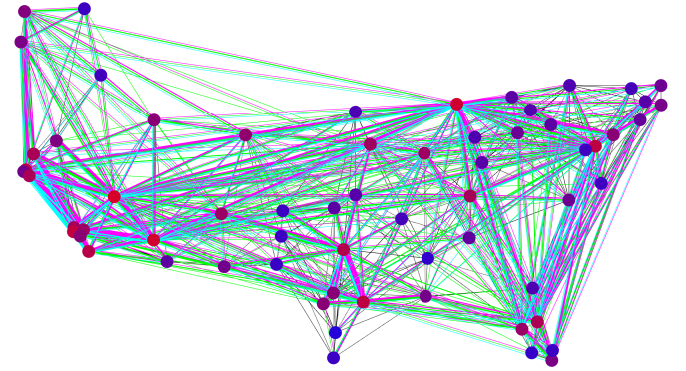}
   \caption{One of the limit cycle states for the simulation with $\delta_i(k) = 10$. %,  for every  $k \in \{0,1,2\}$, $ i \in [n]$. 
   All cities become at least partially infected, depicted by the redish purple color, following \eqref{eq:color}.}
    \label{fig:infected}
\end{figure}

For this simulation, we employ the same parameters as the simulations in Figure \ref{fig:avg} 
% as the first simulation 
except 
$\delta = 10$.
% $\delta_i(k) = 10$,  for every  $k \in \{0,1,2\}$, $ i \in [n]$. 
This system has $\rho(M_{k+3:k}) = 1.4015$ and converges to a limit cycle with three states. This limit-cycle behavior is illustrated via the average infection level plotted in Figure \ref{fig:avg_infected}. 
% \begin{align*}
% &\begin{bmatrix}0.628&	0.628 & 0.628\end{bmatrix} \\
% & \begin{bmatrix} 0.637	&0.637	&0.637 \end{bmatrix}\\
% & \begin{bmatrix}0.649 &	0.649&	0.649 \end{bmatrix}.
% \end{align*}
The values of the three limit cycle states are quite close to each other therefore we only plot one in Figure \ref{fig:infected}.
% \\ 
Simulations show that the limit cycle is independent of the initial condition, given that $x(0) \neq 0$. % as long as it is not the DFE.
%\seb{
This finding implies that if, when factoring in the network connections, the infection rates dominate the healing rates sufficiently, the virus can pervade the network. Therefore, intervention is essential.
%}

\begin{figure}
  \centering
    \begin{overpic}[trim=10 0 10 12,clip,width = \columnwidth] {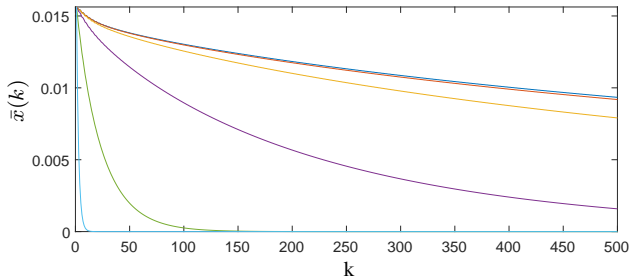}
            \put(51.5,-2){{\parbox{\linewidth}{%
                \footnotesize 
                k}}}
            \put(1,23.25){{\parbox{\linewidth}{%
                    \rotatebox{90}{\footnotesize $\bar{x}(k)$}}}}
        \end{overpic}%\includegraphics[trim=10 0 10 12,clip,width = \columnwidth] {usa_control.eps}
   \caption{Average infection level of the cities over time with the healing rates set to \eqref{eq:local;control} with $\gamma_i = \{0, 0.01, 0.1, 1.0, 10, 100 \}$, for every  $ i \in [n]$. %Blue is for the system with $\rho(M_{k+3:k}) = 1.000$ and red is for $\rho(M_{k+3:k}) < 1.000$.
   }
    \label{fig:avg_control}
\end{figure}

\begin{figure}
  \centering
    \begin{overpic}[trim=10 0 10 12,clip,width = \columnwidth] {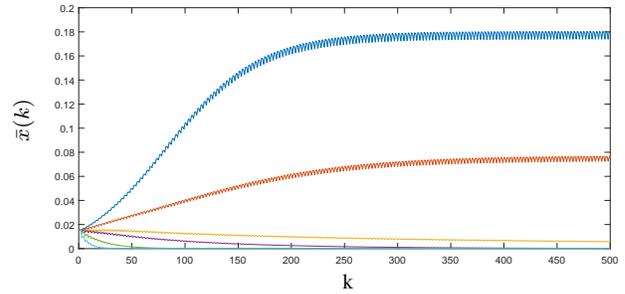}
            \put(51.5,-1){{\parbox{\linewidth}{%
                \footnotesize 
                k}}}
            \put(2,23.25){{\parbox{\linewidth}{%
                    \rotatebox{90}{\footnotesize $\bar{x}(k)$}}}}
        \end{overpic}%\includegraphics[trim=10 0 10 12,clip,width = \columnwidth] {usa_control_ext.eps}
   \caption{Average infection level of the cities over time with the healing rates for the work day period (8:00-16:00) set to \eqref{eq:local;control} with $\gamma_i = \{0, 10, 19.7, 25, 50, 100 \}$ and for the other two periods $\gamma_i=10, \forall  i \in [n]$. %Blue is for the system with $\rho(M_{k+3:k}) = 1.000$ and red is for $\rho(M_{k+3:k}) < 1.000$.
   }
    \label{fig:avg_control_ext}
\end{figure}

In order to see how well the distributed control technique from Section \ref{sec:con} performs, we implement it here. 
In the context of the simulation, the algorithm can be interpreted as a strategy for boosting the healing rates of the more susceptible cities, which could be implemented by deploying mobile treatment clinics, distributing medicine/antidote and installing hand-washing stations in airports and other public places. 
% \seb{
For this simulation we keep the model parameters the same as the previous simulations except we set the healing rates using \eqref{eq:local;control} with
$\gamma_i = \gamma = \{0, 0.01, 0.1, 1.0, 10, 100\}$, for every  $ i \in [n]$,  and $h=0.004$.
Consistent with the results in Theorem~\ref{thm:local:control:ensures:GES} and Corollary~\ref{cor:local:control:ensures:GAS}, the system converges to the DFE in exponential time for nonzero $\gamma$ and in asymptotic time when  $\gamma = 0$. 
% for every $ i \in [n]$. 
Here we explore the effect of $\gamma$ on the convergence rate. 
% Therefore using the same parameters as the rest of the simulations (with no noise) except setting the healing rates using 
% \eqref{eq:local;control} with, for every  $ i \in [n]$, $\gamma_i = \{0, 0.01, 0.1, 1.0, 10, 100\}$ and $h=0.004$. 
The average level of infection for each $\gamma$ value is shown in Figure \ref{fig:avg_control}. We see that for this system $\gamma = 0.01$ behaves very similarly to $\gamma = 0$, while $\gamma \geq 10$ eradicates the virus relatively quickly.
% }
Therefore, if there are enough resources available to boost the healing rates of the cities, the virus can be eradicated. 
However, in certain situations there may not be enough resources to implement such viral-combatant measures during every time step.
Nevertheless, as we saw in the simulations in Figure \ref{fig:avg}, it is not necessary that the healing rates dominate the infection rates at every time step, or even a majority of the time, in order for the virus to be eradicated.
In this set of simulations we explore the effectiveness of the distributed control strategy proposed in Section \ref{sec:con} when the redesign of the healing parameters can only be implemented for part of the period. Given our flight example, the constraint can be interpreted as there only being enough resources to boost the healing rates of the cities during the day, however not in the early morning or at night. 
Therefore, 
% similar to the simulation in Figure \ref{fig:avg_control} 
we implement the controller from \eqref{eq:local;control} but only during the work day (8:00-16:00). 
For the other two periods we set $\delta = 10$ for every city, similar to the simulation in Figure \ref{fig:avg_infected} that displayed the limit-cycle behavior. 
We run a set of simulations with different %homogeneous 
$\gamma$ values in \eqref{eq:local;control}, $\gamma = \{0, 10, 19.7, 25, 50, 100\}$, where $\gamma = \gamma_i$  for every  $ i \in [n]$.  
We plot the average infection level for each simulation in Figure \ref{fig:avg_control_ext}. 
% \phil{It is not clear what we mean by varying $\gamma_i$. Is it that we suppose that $\gamma_i =\gamma$, and then $\gamma$ takes values $0,10,19.7....$?}
As would be expected, a greater $\gamma$ value is needed to eradicate the virus than when actuation is allowed for all three periods. However, even when no control actuation is available for the majority of the periods, the virus can be eradicated for this system if, for the work day period, 
% the homogeneous 
$\gamma \geq 19.7$. 
% for all $i \in [n]$. 
As would be expected, in the case where the equality holds,  $\rho(M_{k+p:k}) = 1$. 
% for all $k \in \{0,1,2\}$.
% }
This result gives hope that, even when there are constraints on the distributed control strategy, the virus can be eradicated.

% \begin{figure}
%   \centering
%     \includegraphics[trim=10 0 10 12,clip,width = \columnwidth] {usa_control.eps}
%   \caption{Average infection level of the cities over time with the healing rates set to \eqref{eq:local;control} with $\gamma = \{0, 0.01, 0.1, 1.0, 10, 100 \}$. %Blue is for the system with $\rho(M_{k+3:k}) = 1.000$ and red is for $\rho(M_{k+3:k}) < 1.000$.
%   }
%     \label{fig:avg_control}
% \end{figure}
% A question, not addressed within the scope of our main results, is the following: given that there are two networks, one of which is in the healthy state whereas the other is  in the infected state, how intense can the interaction be between the healthy agents and the infected ones before they too get infected? Some simulation results seeking to address this question are presented in Figure~\ref{fig:henrikquestion}. The two subpopulations have the adjacency matrices given by $A_3$, for all $k\geq 0$, and every third time step the link connecting the two subpopulations is set to some value $\epsilon$. The healing rate for the healthy population is set to $10$ and the other is $0.1$.
% The simulations show that, for a small edge weight connecting the two populations, the healthy population remains uninfected. However, as the edge weight gets larger the healthy population slowly becomes infected.
% \begin{figure}
%     \centering
%   \includegraphics[width = \columnwidth] {dt_eps.eps}
%     \caption{Interaction between healthy networks and sick networks.}   \label{fig:henrikquestion}
% \end{figure}

\section{Conclusion}\label{sect:conclusion}
%This paper, for 
Considering  discrete-time periodic  time-varying networks with a  mutating virus, this paper has dealt with the problem of designing a control strategy that  ensures exponential (resp. asymptotic) convergence to the healthy state. Our approach was the following: we first provided  conditions for exponential (resp. asymptotic) convergence to the DFE.  Thereafter, we exploited the proven conditions for the design of a distributed control strategy.

Note that  we have restricted our attention to periodic time-varying systems. Hence, a line of future research could be to remove the periodicity assumption. 
Secondly, the present paper operated under the assumption that there was a \emph{single} virus that was infecting the population. Generalizing this setup to account for \emph{multiple} -- not necessarily two -- competing viruses could be an appealing line of investigation.  
Thirdly, the present paper dealt with a \emph{deterministic} model; an inherent drawback with  deterministic models is  that they do not account for the possibility that the system dynamics can 
% -- as illustrated  in Section~\ref{sect:simulations} -- 
be corrupted by noise. Consequently, deterministic models limit our understanding of the potential behaviors involved and a future 
% work, it could be worthwhile to look at 
direction is to study the stochastic version of the periodic SIS model. Finally,  our simulations indicate, under the assumption that the condition in Theorem~\ref{Charzn:DFE} is violated, the possible existence of a 
limit-cycle behavior. However, a rigorous proof (or a counterexample) for this conjecture remains missing.% and  is also the focus of an ongoing investigation.

\bibliography{ReferencesKTH-Phil}

\section*{Appendix}
\subsection*{Proof of Theorem~\ref{result1:GES}:}
\vspace{-1ex}
We use the cyclic reformulation of a linear periodic system; see {\cite[Section~6.3]{bittanti2009periodic}}.  Specifically, define 
\vspace{-1.5ex}
\begin{equation} \label{eq:tildeM} 
\tilde{M} = \begin{bmatrix}
0 &0& \cdots & 0 & M(p-1)\\
M(0)&0&\cdots &0&0\\
0&M(1)&\cdots&0&0\\
\vdots & \vdots & \ddots & \vdots & \vdots\\
0&0&\cdots&M(p-2) &0
\end{bmatrix}.
\end{equation} 
\vspace{-4.5mm}
Note that 
\begin{equation} \label{eq:tildeMp} 
\tilde{M}^{p} = \begin{bmatrix}
 M_{p:0} 
  &0
 &\cdots 
 & 
%  0 &
 0\\
0& M_{p+1:1}&\cdots &0\\
% 0&0&\cdots&0&0\\
 \vdots 
% & 
% \vdots 
& \ddots & 
 \ddots  &  \vdots
\\
0
% &0
& %\cdots
 &0 
&M_{2p-1:p-1}
\end{bmatrix}.
\end{equation}
\vspace{-2ex}

\noindent
Since $\tilde{M}^{p}$ is a block diagonal matrix, the eigenvalues of $\tilde{M}^{p}$ are the eigenvalues of $M_{p:0}$, $M_{p+1:1}$, $\hdots$, $M_{2p-1:p-1}$. By assumption, $\rho(\mathcal{M}) < 1$. Hence, from \eqref{jsr}  it follows that, for all $k \in [p]^-$, $\rho(M_{p:k}) < 1$, %, $\rho(M_{p+1:1}) < 1$, $\hdots$, $\rho(M_{2p-1:p-1}) < 1$,
%\phil{[Doesn't Prop 4 say these are equivalent now?]}
and therefore, $\rho(\tilde{M}^{p}) < 1$. Since the eigenvalues of $\tilde{M}$ are the $p^{th}$-roots of eigenvalues of $\tilde{M}^{p}$, it follows that $\rho(\tilde{M}) < 1$.
% \\

Since, by Assumptions~\ref{assm:2}--\ref{assm:3}, $M(k)$ is nonnegative, it follows that $\tilde{M}$ is also nonnegative. Therefore, from Lemma~\ref{lem:rantzer}, there exists a  diagonal matrix $Q_{1} \succ 0$ such that $\tilde{M}^\top Q_{1}\tilde{M}-Q_{1}\prec 0$. Let the diagonal blocks of $Q_1$ be denoted by $[Q_1]_k \in \mathbb{R}^{N \times N}$, for all $k \in [p]$. By defining $P_1(k) = [Q_1]_{k+1}$, for all $k \in [p]^-$, it is immediate that $M(k)^\top P_{1}(k+1)M(k) - P_{1}(k) \prec 0$ for all $k \in [p]^-$.
% \\
% \noindent 

Consider the following Lyapunov  function $V_{1}(k, x) = x^\top P_{1}(k)x$. Since $Q_1 \succ 0$ and diagonal, each of the blocks along its diagonal must be positive definite. This implies that, for all $k \in [p]^-$ and for $x \neq 0$, $x^\top P_{1}(k)x > 0$, and hence $V_{1}(k,x) > 0$. Since for all $k \in [p]^-$ $P_{1}(k)$ is positive definite, each eigenvalue of $P_{1}(k)$ is real and positive. Then, since $P_{1}(k)$ is also symmetric, by applying the Rayliegh-Ritz Theorem \cite{horn2012matrix}, we obtain
\vspace{-1ex}
\begin{align}
\lambda_{\min}(P_{1}(k))I \leq P_{1}(k) \leq \lambda_{\max}(P_{1}(k))I, \nonumber
\end{align}
\vspace{-4ex}

\noindent
and hence
\vspace{-1ex}
\begin{align}
\lambda_{\min}(P_{1}(k))\left\|x\right\|^{2} \leq V_{1}(k,x) \leq \lambda_{\max}(P_{1}(k))\left\|x\right\|^{2}.
\end{align}
\vspace{-3ex}

\noindent
Define $\sigma_{1}: = \min\limits_{k \in [p]^-}\lambda_{\min}(P_{1}(k))$, and 
\vspace{-1.5ex}
\begin{equation}\label{eq:sigma2}
    \sigma_{2}: = \max\limits_{k \in [p]^-}\lambda_{\max}(P_{1}(k)).
\end{equation} Since for  $k \in [p]^-$ $\lambda_{\min}(P_{1}(k) >0$ and $\lambda_{\max}(P_{1}(k) >0$, it follows that $\sigma_{1} >0$ and $\sigma_{2} >0$. Thus, we have found positive constants $\sigma_{1}$, $\sigma_{2}$ such that for all $k \in [p]^-$ 
\begin{align}
\sigma_{1}\left\|x\right\|^{2} \leq V_{1}(k,x) \leq \sigma_{2}\left\|x\right\|^{2}.
\end{align}

Define $\Delta V_{1}(k,x) = V_{1}(x(k+1)) - V_{1}(x(k))$. For $x \neq 0$, and for all $k \in [p]^-$, one obtains the following:
\begin{align} \label{deltaV}
% \tiny
\Delta V_{1}(k,x) &= x^\top \hat{M}^\top(k) P_{1}(k+1)\hat{M}(k) - x^\top P_{1}(k)x \nonumber\\
&= x^\top(M^\top(k)P_{1}(k+1)M(k)-P_{1}(k))x \nonumber \\
 &\ \ \ \ -2hx^\top \bar{B}^\top(k)X(k)P_{1}(k+1)M(k)x   \nonumber\\ 
&\ \ \ \ + h^{2}x^\top \bar{B}^\top(k)X(k)P_{1}(k+1)X\bar{B}(k)x  .
\end{align}
From Assumptions~\ref{assm:2}--\ref{assm:3} and Lemma~\ref{lem:eqm}, the following is satisfied:
\vspace{-3ex}
\begin{align}
x^\top( h^{2}x^\top \bar{B}^\top(k)X(k)P_{1}(k+1)X(k)\bar{B}(k) \nonumber \\
 -2hx^\top \bar{B}^\top(k)X(k)P_{1}(k+1)M(k))x \leq 0 . \nonumber
\end{align}
\vspace{-4ex}

\noindent
Hence, from \eqref{deltaV} we obtain $$\Delta V_{1}(k,x) \leq x^\top(M^\top(k)P_{1}(k+1)M(k)-P_{1}(k))x.$$ 
Recall that, for all $k \in [p]^-$, $M^\top(k)P_{1}(k+1)M(k)-P_{1}(k)$ is negative definite,  and therefore $M^\top(k)P_{1}(k+1)M(k)-P_{1}(k)$ is symmetric and each of its eigenvalue is real and negative. Hence, applying Rayleigh-Ritz Theorem yields: for all $k \in [p]^-$
\vspace{-1.5ex}
\begin{align}
\Delta V_{1}(k,x) \leq \lambda_{\max}(M^\top(k)P_{1}(k+1)M(k)-P_{1}(k))\left\|x\right\|^{2} \nonumber.
\end{align}
Since $M^\top(k)P_{1}(k+1)M(k)-P_{1}(k)$ is negative definite, $P_{1}(k)-M^\top(k)P_{1}(k+1)M(k)$ is positive definite, and hence we obtain $\lambda_{\max}(M^\top(k)P_{1}(k+1)M(k)-P_{1}(k)) = -  \lambda_{\min}(P_{1}(k)-M^\top(k)P_{1}(k+1)M(k))$, which leads to
\begin{align} \label{sigma}
\Delta V_{1}(k,x) \leq  -  \lambda_{\min}(P_{1}(k)-M^\top(k)P_{1}(k+1)M(k))\left\|x\right\|^{2} .
\end{align}
\vspace{-4ex}

\noindent
Defining  
\vspace{-2ex} 
\begin{equation}\label{eq:sigma3}
    \sigma_{3}: = \max\limits_{k \in [p]^-}\lambda_{\min}(P_{1}(k)-M^\top(k)P_{1}(k+1)M(k)),
\end{equation} 
\vspace{-2.5ex}

\noindent
it follows from~\eqref{sigma} that $\Delta V_{1}(k,x) \leq -\sigma_{3}\left\|x\right\|^{2}$. Since for all $k \in [p]^-$ $P_{1}(k)-M^\top(k)P_{1}(k+1)M(k)$ is positive definite, it follows that $\sigma_{3} >0$.
% \\

Thus, there exists positive constants, $\sigma_{1}$, $\sigma_{2}$ and $\sigma_{3}$, such that for $x \neq 0$ and for all $k \in [p]^-$,
\vspace{-1ex}
\begin{align}
\sigma_{1}\left\|x\right\|^{2} \leq V_{1}(k,x) \leq \sigma_{2}\left\|x\right\|^{2} \label{a1}\\
\Delta V_{1}(k,x) \leq  -  \sigma_{3}\left\|x\right\|^{2}. \label{a2}
\end{align}
\vspace{-3ex}

\noindent
By Assumption \ref{assm:1}, %the $p$-periodicity assumption, 
$M(k+p) = M(k)$ for every $k \in \mathbb{Z}_{\geq 0}$. Hence, over every successive interval of size $p$,  the matrix $\tilde{M}$ remains the same. This implies that $P_{1}(k+p) = P_{1}(k)$ for every $k \in \mathbb{Z}_{\geq 0}$. Hence, we can use the \emph{same} Lyapunov function over \emph{every} successive interval of size $p$. Thus, repeating the same analysis as in the interval $[0, p-1]$ for every successive interval of size $p$ results in inequalities~\eqref{a1} and \eqref{a2} being satisfied for all  $k \in \mathbb{Z}_{\geq 0}$ and for all $x \in [0,1]^n$.  Therefore, from Lemma~\ref{thm:vidyasagar:GES},   the system converges exponentially fast to the  DFE, for all $x(0) \in [0, 1]^n$.~$\square$

%\subsubsection*{Rate of convergence to the DFE}

Next, we explore the rate of convergence to the DFE. 
%\phil{[it's not clear to me why this is a remark. I feel like it should be part of Theorem 1 or at least a corollary (or a proposition), especially given the length of the proof]}

\begin{prop}[Rate of Convergence] \label{rem:decay:rate}
Under the assumptions of Theorem~\ref{result1:GES}, the rate of convergence to the DFE is upper bounded by an exponential with rate $\sqrt{1-\frac{\sigma_3}{\sigma_2}}$, where $\sigma_2$ and $\sigma_3$ are defined in \eqref{eq:sigma2} and \eqref{eq:sigma3}, respectively.~$\blacksquare$ 
\end{prop}
\textit{Proof:} The expression for the rate, $\sqrt{1-\frac{\sigma_3}{\sigma_2}}$, follows directly from \eqref{a1}-\eqref{a2} and \cite[Theorem~23.3]{rugh1996linear}.
Now we show that the rate is well-defined, that is, $0 \leq \sqrt{1-\frac{\sigma_3}{\sigma_2}} <1$. To see this, consider the following: notice that, since $\sigma_2 > 0$ and $\sigma_3 > 0$, it suffices to show that $\sigma_2 \geq \sigma_3$. Towards this end, observe that, for all $k \in [p]^-$, both $P_{1}(k)$ and $M^\top(k)P_{1}(k+1)M(k)$ are symmetric. Applying Weyl's inequalities {\cite[Corollary 4.3.15]{horn2012matrix}} to $P_{1}(k) - M^\top(k)P_{1}(k+1)M(k)$, one obtains, for all $k \in [p]^-$ and $i \in [n]$:
% \begin{equation*}
% \begin{split} 
\vspace{-2ex}
\begin{multline*}
    \lambda_{i}(P_{1}(k) - M^\top(k)P_{1}(k+1)M(k)) \leq \\
\lambda_{i}(P_{1}(k)) + %\nonumber  \\ 
\lambda_{\max}(- M^\top(k)P_{1}(k+1)M(k)), \nonumber
\end{multline*}
\vspace{-1ex}
% \end{split}
% \end{equation*}
which implies 
\vspace{-0ex}
\begin{align}
\begin{split}\lambda_{\max}(P_1(k)) & \geq \lambda_{\max}(P_1(k)- M^\top(k)P_{1}(k+1)M(k)) \label{ineq:1} \end{split} \\
&  \geq \lambda_{\min}(P_1(k)- M^\top(k)P_{1}(k+1)M(k))  \label{ineq:2} 
\end{align}
% \begin{equation}
%   \implies  \sigma_2 \geq \sigma_3, \label{ineq:3}  
% \end{equation}
\vspace{-3ex}

\noindent
where 
% inequality~
\eqref{ineq:1} holds because $- M^\top(k)P_{1}(k+1)M(k)$ is negative semidefinite, implying
% which implies 
$\lambda_{\max}(- M^\top(k)P_{1}(k+1)M(k)) \leq 0$. %Inequality
% ~\eqref{ineq:2} follows from the positive definiteness of $P_{1}(k) - M^\top(k)P_{1}(k+1)M(k)$, \phil{[doesn't this follow from the definition of max and min? I.e. the min is less than or equal to the max regardless of the sign of the values]} and, finally, 
%\eqref{ineq:3} 
% is a consequence of two things: 
By definitions of $\sigma_2$ and $\sigma_3$, and since \eqref{ineq:2} is satisfied for all $k \in [p]^-$, it follows that $\sigma_2 \geq \sigma_3$.~$\square$
%\phil{[the structure is a little weird at the end here. It's not clear that (25) should come in the same align environment as (23)-(24)... especially since the step from (24) to (25) is explained in a separate sentence that is long.]}

\vspace{-2ex}
\subsection*{Proof of Proposition~\ref{equal:1:GAS}:}
\vspace{-1ex}
By assumption, $\rho(\mathcal{M}) = 1$. 
This implies, from the definition of joint spectral radius,  that, for all $k \in [p]^-$, $\rho(M_{p:k}) \leq 1$,  and therefore, $\rho(\tilde{M}^{p}) \leq 1$. Since the eigenvalues of $\tilde{M}$ are the $p^{th}$-roots of eigenvalues of $\tilde{M}^{p}$, it follows that $\rho(\tilde{M}) \leq 1$. For the case where $\rho(\tilde{M}) < 1$, from the proof of Theorem~\ref{result1:GES}, the DFE of \eqref{eq:matrixform} is GAS. Hence, in the rest of the proof, we focus on the case where $\rho(\tilde{M}) = 1$.
% \\}

Suppose that  $\rho(\tilde{M}) = 1$. \seb{Since by Assumptions~\ref{assm:2}--\ref{assm:5}, $M(k)$, for each $k \in [p]^{-}$, is irreducible and nonnegative, it %also
follows that $\tilde{M}$ is also irreducible and nonnegative}. Hence, from Lemma~\ref{lem:pare}, there exists a positive diagonal matrix $Q_2$ such that $\tilde{M}^\top Q_{2}\tilde{M}-Q_{2}\preccurlyeq 0$. By defining, for all $k \in [p]^-$, $P_2(k) = [Q_2]_{k+1}$ it is immediate that $M(k)^\top P_{2}(k+1)M(k) - P_{2}(k) \preccurlyeq 0$, where $k \in [p]^-$.\\
Consider the  Lyapunov candidate  function $V_{2}(k, x) = x^\top P_{2}(k)x$. Observe that, by analogous reasoning as in proof of Theorem~\ref{result1:GES}, for all $k \in [p]^-$ and for $x \neq 0$, $V_{2}(k, x) > 0$. Define $\Delta V_{2}(k,x) = V_{2}(x(k+1)) - V_{2}(x(k))$. For $x \neq 0$, from~\eqref{eq:matrixform}, one obtains:
\vspace{-1ex}
\begin{align*}
\Delta V_{2}(k,x) &= x^\top \hat{M}^\top(k) P_{2}(k+1)\hat{M}(k)x(k) - x^\top P_{2}(k)x\\
&= x^\top(M^\top(k)P_{2}(k+1)M(k)-P_{2}(k))x \\ 
& \ \ \ \  -2hx^\top \bar{B}^\top(k)X(k)P_{2}(k+1)M(k)x(k)  \\ 
& \ \ \ \ + h^{2}x^\top \bar{B}^\top(k)X(k)P_{2}(k+1)X(k)\bar{B}(k)x \\
&\leq h^{2}x^\top \bar{B}^\top(k)X(k)P_{2}(k+1)X(k)\bar{B}(k))x \\ 
& \ \ \ \  -2h x^\top \bar{B}^\top(k) X(k)P_{2}(k+1)M(k)x  \\ 
&=  h^{2}x^\top \bar{B}^\top(k)X(k)P_{2}(k+1)X(k)\bar{B}(k)x \\ 
& \ \ \ \  - h x^\top \bar{B}^\top(k)X(k)P_{2}(k+1)M(k)x\\
& \ \ \ \ -h^{2} x^\top \bar{B}^\top(k)X(k)P_{2}(k+1)\bar{B}(k)x \\
&  \ \ \ \  -h x^\top \bar{B}^\top(k)X(k)P_{2}(k+1)(I-hD(k))x \\ 
& \leq  h^{2} x^\top \bar{B}^\top (k) X(k) P_{2}(k+1)X(k)\bar{B}(k)x  \\ 
& \ \ \ \ -hx^\top \bar{B}^\top (k)X(k)P_{2}(k+1)M(k)x \\ 
& -h^{2}x^\top \bar{B}^\top (k)X(k)P_{2}(k+1)\bar{B}(k)x \\ 
& \leq -h^{2}x^\top \bar{B}^\top (k) X(k) P_{2}(k+1)(I-X(k))\bar{B}(k)x \\ 
&   \ \ \ \ - hx^\top \bar{B}^\top (k)X(k)P_{2}(k+1)M(k)x  \\ 
& \leq - hx^\top \bar{B}^\top (k)X(k)P_{2}(k+1)M(k)x  \\ 
& \leq \hspace{2mm} 0. %\nonumber
\end{align*}

\vspace{-2ex}

It can be immediately seen that if $x=0$, then for all $k \in [p]^-$, $\Delta V_{2}(k,x) =0$. For every $k \in \mathbb{Z}_{\geq 0}$, by Assumptions~\ref{assm:2} and~\ref{assm:5}, $\bar{B}(k)$ (and hence $M(k)$) is nonzero and nonnegative, whereas, from Lemma~\ref{lem:pare}, $P_{2}(k)$ is a positive diagonal matrix. Hence, if, for all $k \in [p]^-$,  $- hx^\top \bar{B}^\top (k)X(k)P_{2}(k+1)M(k)x =0$  then $x=0$.

%{\color{red}I don't follow this next sentence...}
For reasons, similar to those outlined in the proof of Theorem~\ref{result1:GES},  the aforesaid analysis can be repeated over \emph{every} successive interval of size $p$, which yields $V_{2}(k,x) >0$ and $\Delta V_{2}(k,x) \leq 0$ for every $k \in \mathbb{Z}_{\geq 0}$.  Moreover, it can be immediately seen that $V_{2}(k,x)$ is radially unbounded, since $V_{2}(k,x) = \left\| P_{2}(k)^{\frac{1}{2}}x \right\|^{2}$. Therefore, from Lemma~\ref{prop:vidyasagar},   the DFE is GAS.~$\square$

\vspace{-2ex}

\begin{table*}[bp]
 %\hline
\hrule
    \centering \normalsize
\seb{\begin{equation}
 \bar{\bar{M}}(X) :=  
\begin{bmatrix}
\hat{M}_{p-1}(pq)\cdots\hat{M}_{0}(pq)&0& \cdots %&\cdots
&0\\
0&\hat{M}_p(pq)\cdots\hat{M}_1(pq)&\cdots &0\\
\vdots&\vdots&\ddots%&\cdots
&0\\
0&0&\cdots%&\cdots
&\hat{M}_{2p-2}(pq)\cdots\hat{M}_{p-1}(pq)
\end{bmatrix} \nonumber
\end{equation}}
    \caption{\seb{Definition of $\bar{\bar{M}}(X)$.}}
    \label{tab:mdoublebar}
\end{table*}

\subsection*{Proof of Proposition~\ref{prop:nece:global:AS}:}
\seb{%It turns out that
System~\eqref{eq:matrixform} can be represented as a nonlinear \emph{time-invariant} system, using the technique provided in \cite[Section~\rom{4}.C]{ye2018evolution}. Towards this end, we first write~\eqref{eq:matrixform} using \eqref{eq:Mhat} as:
\begin{equation} \label{eq:state:rewritten}
    x(pq+k+1) =\hat{M}(pq+k)x(pq+k),
\end{equation}
where $p$ is the periodicity of system~\eqref{eq:matrixform}, and $q$ is any nonnegative integer.
Recall from~\eqref{eq:Mhat} that 
\begin{align}
\hat{M}(pq+k) &= M(pq+k)-hX(pq+k)\bar{B}(pq+k)  \nonumber \\ \label{periodicity:expolit}
\ \ \ & =M(k)-hX(pq+k)\bar{B}(k),
%\ \ \ & = \hat{M}_k(pq+k) %\label{hat:new:notation}
\end{align}
where~\eqref{periodicity:expolit} is a consequence of Assumption~\ref{assm:1}. 
Define 
\begin{equation}\label{eq:mhatk}
    \hat{M}_k(pq) := M(k \text{ mod }p)-hX(pq+k)\bar{B}(k \text{ mod }p).
\end{equation} %Hence, %for all $k \in [p]^-$ and for $q =0,1,2, \hdots$, we have $\hat{M}(pq+k) =\hat{M}_k(pq)$.

By concatenating the state vector over an interval of size $p$, we define a new state variable $y \in \mathbb{R}^{pn}$, \vspace{-1ex}
\begin{equation}\label{TI:2}
    y(pq) := \begin{bmatrix} 
    y_1(pq) \\
    y_2(pq)\\
    \vdots \\
    y_p(pq)
    \end{bmatrix} =
    % \quad
    \begin{bmatrix} 
    x(pq) \\
    x(pq+1)\\
    \vdots\\
    x(pq+p-1)
    \end{bmatrix}.
\end{equation}
We are interested in studying how $y(pq)$ evolves for $q \in \{0,1,2, \hdots\}$. From~\eqref{TI:2}, we have \vspace{-1ex}
 \begin{equation}\label{TI:3}
    y(p(q+1)) = 
    \begin{bmatrix} 
     y_1(p(q+1)) \\
     y_2(p(q+1))\\
     \vdots\\
     y_p(p(q+1))
     \end{bmatrix} =
     %\quad
     \begin{bmatrix} 
     x(p(q+1)) \\
     x(p(q+1)+1)\\
     \vdots\\
    x(p(q+1)+p-1)
     \end{bmatrix}. \nonumber
\end{equation}
From~\eqref{eq:state:rewritten}, \eqref{periodicity:expolit}, and \eqref{eq:mhatk}, we know that \vspace{-1ex}
%\tiny
\begin{align*} 
x(p(q+1)) &=  \hat{M}_{p-1}(pq)\cdots\hat{M}_{0}(pq)x(pq) \nonumber \\
x(p(q+1)+1) &=  \hat{M}_p(pq)\cdots\hat{M}_1(pq)x(pq+1) %\ \ \ \ \ \ \ \ \ \ 
\nonumber \\
&\ \; \vdots 
\\
% \end{align*}
% \begin{align*} 
%  \ \ \ \ \ 
 x(p(q+1)+p-1) &=  \hat{M}_{2p-2}(pq)\cdots\hat{M}_{p-1}(pq)x(pq+p-1). \nonumber 
\end{align*}%\normalsize
Hence, we can rewrite %~\eqref{TI:3}
the dynamics 
as% a discrete-time nonlinear time-invariant system:
\vspace{-1ex}
\begin{equation} \label{eq:TI:final}
    y(p(q+1)) = \bar{\bar{M}}(X)y(pq),
\end{equation} 
\vspace{-4ex}

\noindent
where $\bar{\bar{M}}(X)$ is defined in Table \ref{tab:mdoublebar}. %\philnew{TBD: could we display the matrix $\bar{\bar{M}}(X)$ better?}\\
Note that as a consequence of~\eqref{eq:mhatk}, the dynamics of the system in~\eqref{eq:TI:final}, for any $q\in \mathbb{Z}_+$, depend only 
on  the matrices $M(k)$ and $\bar{B}(k)$, for all $k \in [p]^-$, and the
state vector. % as a consequence of the modulo function. 
% which implies that~
Therefore, \eqref{eq:TI:final} is a discrete-time nonlinear time-invariant system.
% 
% Rewriting $\bar{\bar{M}}(X)$ using~\eqref{TI:11} and~\eqref{TI:12}, and l
Linearizing~\eqref{eq:TI:final} around %the DFE, i.e., $x=0$,
$y = 0$ yields the following:
\vspace{-1ex}
\begin{equation}
 \label{eq:matrixform:linearized}
y(p(q+1)) = \tilde{M}^py(pq),
\end{equation}
where $\tilde{M}^p$ is defined in \eqref{eq:tildeMp}.%\\

By way of contraposition, assume 
that, for all $k \in [p]^-$,  $\rho(M_{k +p:k}) > 1$.
%By the assumption that %there exists  $k^\prime \in [p]^-$ such that\\
%$\rho(M_{k+p:k}) > 1$.
This assumption %(see \eqref{eq:tildeMp}), 
implies that  $\rho(\tilde{M}^p) >1$, since $\tilde{M}^p$ is a block diagonal matrix. %Since eigenvalues of $\tilde{M}$ are $p^{th}$-roots of eigenvalues of $\tilde{M}^p$,  it follows that $\rho(\tilde{M}) >1$.
Therefore, from Proposition~\ref{thm:vidyasagar:unstable}, $y=0$ is an unstable equilibrium of system~\eqref{eq:TI:final}.
% Therefore 
%From~\eqref{TI:2} it is immediate that $y =0$
Since, by~\eqref{TI:2}, $y=0$ corresponds to the DFE of~\eqref{eq:matrixform}, %and. therefore, 
%Thus, since the system in \eqref{TI:2} is equivalent to the one in \eqref{eq:matrixform},
the DFE is an unstable equilibrium of system~\eqref{eq:matrixform}.~$\square$
}

\vspace{-2ex}
\subsection*{Proof of Theorem~\ref{thm:local:control:ensures:GES}:}
\vspace{-1ex}
 Consider healing rates as in \eqref{eq:local;control}. Define $\hat{B}(k) = \diag(\sum\limits_{j=1}^{n}\bar{\beta}_{ij}(k) + \gamma_{i})$. Then substituting \eqref{eq:local;control} into~\eqref{eq:dt}, and rewriting the system in matrix form yields:
\begin{equation} \label{eg:new:system}
x(k+1) = x(k) + h ((I-X(k))\bar{B}(k)-\hat{B}(k))x(k).
\end{equation}
Define $M^1(k)= I - h\hat{B}(k) + h\bar{B}(k)$. As a consequence of Assumptions~\ref{assm:2} %and~\ref{assm:3},
\seb{and since, for each $i \in [n]$, $\gamma_i$ is chosen such that $h\sum\limits_{j=1}^{n}\bar{\beta}_{ij}(k) + h\gamma_{i} \leq 1$,} for all $k \in [p]^-$, $M^1(k)$ is nonnegative. Moreover, since \seb{$h>0$, and} $\gamma_{i} >0$ $\forall i \in [n]$,  each row $i$ of $M^1(k)$ satisfies $\sum\limits_{j=1}^{n} [M^1(k)]_{ij} <1$ for all $k \in [p]^-$. %\\

By building  $\tilde{M}^1$ analogous to $\tilde{M}$ in \eqref{eq:tildeM}, the structure of $\tilde{M}^1$ immediately gives that $\tilde{M}^1$ is nonnegative  and that each row satisfies $\sum\limits_{j=1}^{n} [\tilde{M}^1]_{ij} <1$. Thus by definition of the infinity norm of a matrix, $\left\|\tilde{M}^1\right\|_{\infty} <1$, which, due to \cite[Theorem~5.6.9]{horn2012matrix}, further implies that $\rho(\tilde{M}^1) < 1$. From the proof of Theorem~\ref{result1:GES}, it is clear that 
 $\rho(\tilde{M}^1) < 1$ ensures that the DFE is GES.~$\square$\\

\end{document}